\documentclass[preprint]{elsarticle}

\ifx\newdefinition\undefined\def\newdefinition#1#2{}\fi

\ifx\mainDir\undefined
  \def\mainDir{.}
  \IfFileExists{\mainDir/tex-header}{}{\def\mainDir{..}}
  \IfFileExists{\mainDir/tex-header}{}{\def\mainDir{...}}
  \IfFileExists{\mainDir/tex-header}{}{\def\mainDir{....}}
  \IfFileExists{\mainDir/tex-header}{}{\def\mainDir{.....}}
  \IfFileExists{\mainDir/tex-header}{}{\def\mainDir{......}}
\fi
\IfFileExists{\mainDir/tex-header.tex}{}{\PackageError{tex-header}{Could not auto-detect the project directory.}{You should include the tex-header file using input{./tex-header} or input{../tex-header} or ... or input{....../tex-header}.}}

\ifx\IfCompilationRoot\undefined

\usepackage[T1]{fontenc}
\usepackage{natbib}
\usepackage{standalone}
\usepackage{amsmath}
\usepackage{amssymb}
\usepackage{amsthm}
\usepackage{graphicx}
\usepackage{proof}
\usepackage{float}
\usepackage{color}
\usepackage{soul}
\usepackage{listings}
\usepackage{stmaryrd}
\usepackage{enumitem}
\usepackage{tikz}
\usetikzlibrary{trees,arrows,arrows.meta,shapes,decorations.pathmorphing,backgrounds,positioning,fit,petri}
\usepackage{algorithm}
\usepackage[noend]{algpseudocode}
\usepackage{url}
\usepackage{booktabs} 

\usepackage[font=small]{subcaption}
\usepackage[font=small]{caption}
\usepackage{hyperref}

\usepackage{tkz-graph}

\usepackage{amsmath}
\usepackage{amsthm}

\newtheorem{observation}{Observation}

\newdefinition{example}{Example}
\newdefinition{definition}{Definition}
\newdefinition{remark}{Remark}
\newdefinition{notation}{Notation}
\newdefinition{problem}{Problem}

\newtheorem{proposition}{Proposition}
\newtheorem{theorem}{Theorem}
\newtheorem*{theorem*}{Theorem}
\newtheorem{corollary}{Corollary}
\newtheorem{lemma}{Lemma}

\newtheorem*{claim*}{Claim}

\makeatletter\def\@font@warning#1{}\makeatother
\makeatletter\def\HyPsd@CatcodeWarning#1{}\makeatother

\makeatletter
\newbox\dottedarrow@box
\setbox\dottedarrow@box\hbox
  {%
    \begin{tikzpicture}[>=stealth]
      \draw[blue,thick,dotted,->] (0,0) -- (1.5em,0);
    \end{tikzpicture}%
  }
\newcommand*\dottedarrow
  {\relax\ifmmode\expandafter\dottedarrow@m\else\expandafter\dottedarrow@t\fi}
\newcommand*\dottedarrow@t[1][1.5em]
  {\resizebox{#1}{!}{\raisebox{.5ex}{\usebox\dottedarrow@box}}}
\newcommand*\dottedarrow@m[1][]
  {%
    \if\relax\detokenize{#1}\relax
      \mathchoice
        {\dottedarrow@t}
        {\dottedarrow@t}
        {\dottedarrow@t[1.1em]}
        {\dottedarrow@t[0.9em]}%
    \else
      \dottedarrow@t[#1]%
    \fi
  }
\makeatother

\newcommand{\jhvar}{{\mathcal{E}}}  
\newcommand{\sfunspace}[1]{\jhvar(#1)}  
\newcommand{\jhspace}[1]{\sfunspace{#1}}  

\newcommand{\tuplespace}[2]{ #1^#2 } 

\newcommand{\sfuntuple}{f}

\newcommand{\Dfuntuple}{{\bigmeetp{\jhspace{\Lat}}}}

\newcommand{\dplusfuntuple}{\delta}
\newcommand{\dapprox}{\sigma}

\newcommand{\sfun}[1]{\sfuntuple_{#1}}

\newcommand{\Dfun}[1]{\Dfuntuple{#1}}	

\newcommand{\dplusfun}[1]{\dplusfuntuple_{#1}}
\newcommand{\dapproxfun}[1]{\dapprox_{#1}}

\newcommand{\sfunapp}[2]{\sfun{#1}( {#2} )}

\newcommand{\Dfunapp}[2]{\big(\Dfun{#1}\big)({#2})}	
			
\newcommand{\dplusfunapp}[2]{\dplusfun{#1}( {#2} )}	
\newcommand{\dapproxfunapp}[2]{\dapproxfun{#1}( {#2} )}

\newcommand{\C}{\Con}                   
\newcommand{\Con}{\Lat}					

\newcommand{\Lat}{L}   
\newcommand{\M}{\mbox{\bf M}}   

\newcommand{\bote}{\bot}	   
\newcommand{\tope}{\top}	   

\newcommand{\join}{\sqcup}						
\newcommand{\joinp}[1]{\join_{{\scriptscriptstyle #1}}}
\newcommand{\bigjoin}{\bigsqcup}						
\newcommand{\bigjoinp}[1]{\bigjoin\nolimits_{{\scriptscriptstyle #1}}}

\newcommand{\meet}{\sqcap}						
\newcommand{\meetp}[1]{\meet_{{\scriptscriptstyle #1}}}
\newcommand{\bigmeet}{\bigsqcap}						
\newcommand{\bigmeetp}[1]{\bigmeet\nolimits_{{\scriptscriptstyle #1}}}
	
\newcommand{\latminus}{\ominus}

\newcommand{\cleq}{\sqsubseteq}					
\newcommand{\cgeq}{\sqsupseteq}		
\newcommand{\cl}{\sqsubset}			
\newcommand{\cg}{\sqsupset}			
\newcommand{\lcov}{\prec}			
\newcommand{\gcov}{\succ}			
\newcommand{\lcoveq}{\preceq}			

\newcommand{\hleq}{\cleq_\jhvar}         
\newcommand{\hgeq}{\cgeq_\jhvar} 

\newcommand{\imp}{\rightarrow}

\newcommand{\defsymbol}{\stackrel{{\scriptscriptstyle \textup{\texttt{def}}}}{=}}

\newcommand{\laguerre}[2]{{\mathcal L}_{#1}(#2)}  
\newcommand{\rook}[2]{r{(#1,#2)}} 
\newcommand{\rookp}[2]{r^\prime{(#1,#2)}} 
\newcommand{\Rook}[2]{{\mathcal R}_{#1}(#2)}  

\newcommand{\fres}[2]{#1{\upharpoonright}_{#2}}

\newcommand{\func}[3]{#1 : #2 \to #3}

\newcommand{\downset}[1]{{\downarrow}{#1}}
\newcommand{\covers}[1]{{\downarrow^{\scriptscriptstyle 1}}{{#1}}}

\newcommand{\selfcovers}[1]{{\downarrow^{\scriptstyle 01}}{{#1}}}

\newcommand{\cov}{\prec}    

\newcommand{\varijoin}{\mathcal{J}}
\newcommand{\ijoin}[1]{\varijoin(#1)}
\newcommand{\jdown}[1]{\left\downarrow^{{\scriptscriptstyle\varijoin}} #1\right.}

\newcommand{\sop}{\ominus}

\newcommand{\Cset}{\mathtt{Con}}
\newcommand{\Fset}{\mathtt{Fail}}
\newcommand{\Sset}{\mathtt{Sup}}

\newcommand{\Cseta}[1]{\mathtt{Con}_{#1}}
\newcommand{\Fseta}[1]{\mathtt{Fail}_{#1}}
\newcommand{\Sseta}[1]{\mathtt{Sup}_{#1}}

\newcommand{\fcal}{\mathcal{F}}

\newcommand{\pcal}{\mathcal{P}}

  \makeatletter
  \newcounter{inputDepth}
  \def\IfCompilationRoot@refresh{\ifnum\value{inputDepth}=0\global\def\IfCompilationRoot##1{##1}\else\global\def\IfCompilationRoot##1{}\fi}
  \IfCompilationRoot@refresh{}
  \let\realInput\input
  \global\def\modifiedInput#1{\addtocounter{inputDepth}{1}\IfCompilationRoot@refresh{}\realInput{#1}\addtocounter{inputDepth}{-1}\IfCompilationRoot@refresh{}}
  \let\input\modifiedInput
  \makeatother
\fi

\makeatletter\def\@font@warning#1{}\makeatother

\begin{document}

\begin{frontmatter}
    \title{Counting and Computing Join-Endomorphisms in Lattices (Revisited)\tnoteref{t1}}
    
    \author[3]{Carlos Pinz\'on}
    \ead{carlos.pinzon@inria.fr}
    
    \author[4]{Santiago Quintero}
    \ead{squinter@polytechnique.fr}
    
    \author[5]{Sergio Ram\'irez}
    \ead{ssramirezr@eafit.edu.co}
    
    \author[1]{Camilo Rueda}
    \ead{crueda@javerianacali.edu.co}
    
    \author[1,2]{Frank Valencia\corref{cor1}}
    \ead{frank.valencia@lix.polytechnique.fr}
    
    \address[1]{Departament of Electronics and Computer Science \\ Pontificia Universidad Javeriana, Cali, Colombia}
    
    \address[2]{CNRS, LIX \'Ecole Polytechnique de Paris, France}
    
    \address[3]{INRIA, \'Ecole Polytechnique de Paris, France}
    
    \address[4]{LIX \'Ecole Polytechnique de Paris, France}

    \address[5]{Universidad EAFIT, Medellin, Colombia}
    
    
    \cortext[cor1]{Corresponding author}
    
    \tnotetext[t1]{This work has been partially supported by the ECOS-NORD project FACTS (C19M03)}
    
    \begin{abstract}
    Structures involving a lattice and join-endomorphisms on it are ubiquitous in computer science. We study the cardinality of the set $\sfunspace{\Lat}$ of all join-endomorphisms of a given finite lattice $\Lat$. In particular, we show for $\M_n$, the discrete order of  $n$ elements extended with top and bottom, $| \sfunspace{\M_n} | =n!\laguerre{n}{-1}+(n+1)^2$ where $\laguerre{n}{x}$ is the Laguerre polynomial of degree $n$.
    We also study the following problem: {Given a lattice $L$ of size $n$ and a set $S\subseteq \sfunspace{\Lat}$ of size $m$, find the greatest lower bound $\Dfun{S}$}. The join-endomorphism $\Dfun{S}$ has meaningful interpretations in epistemic logic, distributed systems, and Aumann structures. We show that this problem can be solved with worst-case time complexity in $O(mn)$ for distributive lattices and $O(mn + n^3)$ for arbitrary lattices. In the particular case of modular lattices, we present an adaptation of the latter algorithm that reduces its average time complexity. We provide theoretical and experimental results to support this enhancement. The complexity is expressed in terms of the basic binary lattice operations performed by the algorithm.
\end{abstract}
    
    
\end{frontmatter}    


\section{Introduction}

There is a long established tradition of using lattices to model structural entities in many fields of mathematics and computer science. For example, lattices are used in concurrency theory to represent the hierarchical organization of the information resulting from agent's interactions~\cite{knight:hal-00761116}. \emph{Mathematical morphology} (MM), a well-established theory for the analysis and processing of geometrical structures, is founded  upon lattice theory~\cite{bloch-mm-2007,ronse-mmclat-1990}. 
Lattices are also used as algebraic structures for modal and epistemic logics as well as Aumann structures (e.g., modal algebras and constraint systems~\cite{guzman:hal-01257113}).

In all these and many other applications, lattice  {join-endomorphisms} appear as fundamental. A \emph{join-endomorphism} is a function from a lattice to itself that preserves finite joins.  In MM, join-endomorphisms correspond to one of its fundamental operations; \emph{dilations.} In modal algebra, they correspond via duality to the box modal operator. In epistemic settings, they represent belief or knowledge of agents. In fact, our own interest in lattice theory  derives from using join-endomorphisms to model the perception that agents may have of a statement in a lattice of partial information~\cite{guzman:hal-01257113}.   

For finite lattices, devising suitable algorithms to compute lattice  maps with some given properties would thus be of  great utility.   We are interested in constructing algorithms for computing lattice join-endomorphisms. This requires, first, a careful study of the space of such maps to have a clear idea  of how particular lattice structures impact on the size of the space. We are, moreover, particularly interested in computing the \emph{maximum} join-endomorphism below a given collection of join-endomorphisms. This turns out to be important, among others, in spatial computation (and in epistemic logic) to model the distributed information (resp. distributed knowledge) available to a set of agents as conforming a group~\cite{GuzmanKQRRV19}. It could also be regarded as the maximum perception consistent with (or derivable from)  a collection of perceptions of a group of agents.

{\bf Problem.} Consider the set $\sfunspace{\Lat}$ of all join-endomorphisms of a finite lattice $\Lat$. The set $\sfunspace{\Lat}$ can be made into a lattice by ordering join-endomorphisms point-wise wrt the order of $\Lat$. We investigate the following maximization problem: \emph{Given a lattice $\Lat$ of size $n$ and a set $S\subseteq \sfunspace{\Lat}$ of size $m$, find in $\sfunspace{\Lat}$ the  greatest lower bound of $S$, i.e. $\Dfun{S}.$}  Simply taking $\sigma:\Lat \to \Lat$ with $\sigma(e) \defsymbol \bigmeetp{\Lat}\{ f(e) \mid f \in S \}$ does not solve the problem as $\sigma$ may not be a join-endomorphism. Furthermore, since $\sfunspace{\Lat}$ can be seen as the search space, we also consider the problem of determining its cardinality. Our main results are the following.

{\bf This paper.}  We present characterizations of the exact cardinality of $\sfunspace{\Lat}$ for some fundamental lattices. Our contribution is to establish the cardinality  of $\sfunspace{\Lat}$ for the stereotypical non-distributive lattice $L={\M_n}.$ We show that  $|\sfunspace{\M_n}|$ equals $r_0^n+ \ldots + r_n^n +r_1^{n+1}=n!\laguerre{n}{-1}+(n+1)^2$ where $r^m_k$ is the number of ways to place $k$ non-attacking rooks on an $m\times m$ board and $\laguerre{n}{x}$ is the Laguerre polynomial of degree $n$. 
We also present cardinality results for powerset and linear lattices that are part of the lattice theory folklore: The number of join-endomorphisms is $n^{\log_2 n}$ for powerset lattices of size $n$ and $\binom{2n}{n}$ for linear lattices of size $n+1$. Furthermore, we provide algorithms that, given a lattice $\Lat$ of size $n$ and a set $S\subseteq \sfunspace{\Lat}$ of size $m$,  compute  $\Dfun{S}.$  Our contribution is to show that $\Dfun{S}$ can be computed  with worst-case time complexity in  $O(n + m{\log n})$ for powerset lattices, $O(mn^2)$ for lattices of sets, and $O(nm + n^3)$ for arbitrary lattices.


\section{Background: Join-Endomorphisms and Their Space}
\label{sec:back}

We presuppose basic knowledge of order theory~\cite{davey2002introduction} and use the following notions.

\paragraph{Partially Ordered Sets}
Let $(\Lat, \cleq)$ be a partially ordered set (poset), and let $S \subseteq \Con$. We use $\bigjoinp{\Lat} S$ to denote the least upper bound (or \emph{supremum} or \emph{join}) of $S$ in $\Lat$, if it exists. Dually,  $\bigmeetp{\Lat} S$ is the greatest lower bound (glb) (\emph{infimum} or  \emph{meet}) of $S$ in $\Lat$, if it exists. We shall often omit the index $\Lat$ from $\bigjoinp{\Lat}$ and $\bigmeetp{\Lat}$ when no confusion arises.  As usual, if $S= \{ c,d \}$, $c \sqcup d$ and $c \meet d$ represent $\bigjoin S$ and $\bigmeet S$, respectively. If $\Lat$ has a greatest element (top)  $\tope$, and a least element (bottom) $\bote$, we have $\bigjoin \emptyset = \bote$ and $\bigmeet \emptyset = \tope$. The poset $\C$ is \emph{distributive} iff for every $a,b,c \in \Con$, $a \join ( b \meet c) =  (a \join b) \meet (a \join c)$.

Given $n$, we use $\mathbf{n}$ to denote the poset $\{1,\ldots,n\}$ with the linear order $x \cleq y$ iff $x\leq y$. The poset $\bar{\mathbf{n}}$ is the set $\{1,\ldots,n\}$ with the discrete order $x \cleq y$ iff $x=y$. Given a poset $\Lat$, we use $\Lat_\bot$ for the poset that results from adding a bottom element to $\Lat.$ The poset $\Lat^\top$ is similarly defined.

\paragraph{Lattices}
The poset $\C$ is a \emph{lattice} iff each finite nonempty subset of $\Con$ has a supremum and infimum in $\Con$, and it is a \emph{complete lattice} iff each subset of $\Con$ has a supremum and infimum in $\Con$. The lattice $\mathbf{2}^n$ is the $n$-fold Cartesian product of $\mathbf{2}$ ordered coordinate-wise. We define $\M_n$ as the lattice $(\bar{\mathbf{n}}_\bot)^\top.$ A \emph{lattice of sets}  is a set of sets ordered by inclusion and closed under finite unions and intersections. A \emph{powerset lattice} is a lattice of sets that includes all the subsets of its top element. A lattice $\Lat$ is said to be \emph{modular} if for every $a,b,c \in \Lat$, with $a \cleq b$, the equality $a \join (c \meet b) = (a \join c) \meet b$ holds.

\paragraph{Self-maps}
A \emph{self-map} on $\Con$ is a function $f: \Con \to \Con$. A self-map $f$ is \emph{monotonic} if $a \cleq b$ implies $f(a) \cleq f(b)$. We say that $f$ \emph{preserves} the join of $S \subseteq \Con$  iff $f(\bigjoin S) = \bigjoin \{ f(c)\mid c \in S \}$.

\begin{definition}[Downsets, Covers, Join-irreducibility~\cite{davey2002introduction}]
\label{def:covers}
Let $\Lat$ be a lattice and $a,b \in \Lat$. We say that $a$ {\em covers} $b$, written $b \lcov a$ (or $a \gcov b$), if $a \cg b$ and there is no $c \in \Lat$ such that $a \cg c \cg b$. The \emph{down-set} of $a$ is $\downset{a} \defsymbol \{b \in \Lat \mid b \cleq a\}$ and the set of elements \emph{covered} by $a$ is $\covers{a} \defsymbol \{ b \mid b \cov a\}$. We write $b \lcoveq a$ for $b \lcov a$ or $b = a$ and define the \emph{cover set} of $a$ as $\selfcovers{a} \defsymbol \{b \in \Lat \mid b \lcoveq a\}$.
An element $c \in \Lat$ is said to be \emph{join-irreducible} if $c = a \join b$ implies $c = a$ or $c = b$. The set of all join-irreducible elements of $\Lat$ is $\ijoin{\Lat}$ and $\jdown{c} \defsymbol \downset{c} \cap \ijoin{\Lat}$.
\end{definition}

We shall investigate the set of all  join-endomorphisms of a given lattice ordered point-wise. Notice that every finite lattice is a complete lattice. 

\begin{definition}[Join-endomorphisms and their space]
Let $\Lat$ be a complete lattice.  We say that a self-map is  a  \emph{(lattice) join-endomorphism} iff it preserves the join of every finite
subset of $\Lat$.  Define $\jhspace{\Lat}$ as the set of all join-endomorphisms 
of $\Lat$. Furthermore,  given $f,g \in \jhspace{\Lat}$, define $f  \hleq  g$ iff $f(a) \cleq g(a)$
for every $a\in\Lat.$
\end{definition}

The following are immediate consequences of the above definition. 

\begin{proposition}
\label{prop:monotonicity}
Let $\Lat$ be a complete lattice.  $f \in \jhspace{\Lat}$ iff $f(\bot) =\bot$ and $f(a \join b) = f(a) \join f(b)$ for all $a,b\in\Lat$.
If $f$ is a join-endomorphism of $\Lat$ then $f$  is monotonic.
\end{proposition}

Given a set $S \subseteq \jhspace{\Lat}$, where $\Lat$ is a finite lattice, we are interested in  finding the greatest join-endomorphism in $\jhspace{\Lat}$ below the elements of $S$, i.e. $\Dfun{S}$. Since every finite lattice is also a complete lattice, the existence of $\Dfun{S}$ is guaranteed by the following proposition.

\begin{proposition}[\cite{gratzer-latjoinend-1958}]
\label{h-space:proposition}
If $(\Lat,\cleq)$ is a complete lattice, $(\jhspace{\Lat},\hleq)$ is a complete lattice.
\end{proposition}

In the following sections we study  the cardinality of $\jhspace{\Lat}$ for some fundamental lattices  and  provide efficient algorithms to compute $\Dfun{S}$.


\section{The Size of the Function Space}
\label{sec:size-fs}

The main result of this section is Theorem~\ref{size4m-n:lemma}. It states the size of $\sfunspace{\M_n}$.  Propositions \ref{size4powerset:lemma} and \ref{size4linear:lemma}  state, respectively, the size of $\sfunspace{\Lat}$ for the cases when $\Lat$ is a powerset lattice and when $\Lat$ is a total order. These propositions follow from simple observations and they are part of the lattice theory folklore \cite{birkhoff-lattice-1940,jipsen-ramics-2017,santocanale-words-2019}. We include our original proofs of these propositions in the technical report of this paper \cite{quintero:hal-02422624}.

\subsection{Distributive Lattices}

We begin with lattices isomorphic to $\mathbf{2}^n$.
They include \emph{finite boolean algebras} and \emph{powerset} lattices \cite{davey2002introduction}. The size of these lattices are easy to infer from the observation that the join-preserving functions on them are determined by their action on the lattices' atoms.
 
\begin{proposition}\label{size4powerset:lemma} Suppose that $m\geq0.$ Let $\Lat$ be any lattice isomorphic to the product lattice $\mathbf{2}^m$.  Then $|\sfunspace{\Lat}| = n^{\log_2 n}$ where  $n = {2}^m$ is the size of $\Lat$. 
\end{proposition}

Thus powerset lattices and boolean algebras have a super-polynomial, sub-exponen\-tial number of join-endomorphisms. Nevertheless, linear order lattices allow for an exponential number of join-endomorphisms  given by the \emph{central binomial coefficient}. The following proposition is also easy to prove from the observation that the join-endomorphisms over a linear order are also monotonic functions. In fact, this result appears in~\cite{birkhoff-lattice-1940} and it is well-known among the algebraic theory community~\cite{jipsen-ramics-2017,santocanale-words-2019}.

\begin{proposition}\label{size4linear:lemma} Suppose that $n\geq 0.$ Let $\Lat$ be any lattice isomorphic to the linear order lattice $\mathbf{n}_\bot$. Then  $|\sfunspace{\Lat}| = \binom{2n}{n}$.
\end{proposition}

It is easy to prove that $\frac{4^n}{2\sqrt{n}} \leq \binom{2n}{n} \leq  4^n$ for $n\geq 1.$ Together with Proposition~\ref{size4linear:lemma}, this gives us explicit exponential lower and upper bounds for
$|\sfunspace{\Lat}|$ when $\Lat$ is a linear lattice.

\subsection{Non-distributive Case}
\label{ssec:size-fs:non-dist}

The number of join-endomorphisms for some non-distributive lattices of a given size can be much  bigger than that for those distributive lattices of the same size in the previous section.  We will characterize this number for an archetypal non-distributive lattice  in terms of Laguerre (and rook) polynomials.

\emph{Laguerre polynomials} are solutions to Laguerre's second-order linear differential equation
$xy'' + (1 - x)y' + ny = 0$ where $y'$ and $y''$ are the first and second derivatives of an unknown function $y$ of the variable $x$, and $n$ is a non-negative integer. The Laguerre polynomial of degree $n$ in $x$, $\laguerre{n}{x}$
 is given by the summation $\sum_{k=0}^n \binom{n}{k}\frac{(-1)^k}{k!}x^k.$ 
 
The lattice $\M_n$ is non-distributive for any $n\geq 3$. The size of $\sfunspace{\M_n}$  can be succinctly expressed as follows. 
  
\begin{theorem}\label{size4m-n:lemma}   \( |\sfunspace{\M_n}| = (n+1)^2  +  n! \laguerre{n}{-1}.\)
\end{theorem}

In combinatorics rook polynomials are generating functions of the number of ways to place non-attacking rooks on a board.  A \emph{rook polynomial} (for square boards) $\Rook{n}{x}$ has the form $\sum_{k=0}^{n}x^k \rook{k}{n}$  where the (rook) coefficient $\rook{k}{n}$  represents
the number of ways to place $k$ non-attacking rooks on an $n \times n$ chessboard. For instance, $\rook{0}{n} = 1$,  $\rook{1}{n}=n^2$ and
$\rook{n}{n}=n!$. In general $\rook{k}{n}=\binom{n}{k}^2k!$.

Rook polynomials are related to Laguerre polynomials by the equation  
$\Rook{n}{x}=n! x^{n}\laguerre{n}{-x^{{-1}}}$. Therefore, as a direct consequence of the above theorem, we can also characterize $|\sfunspace{\M_n}|$ in combinatorial terms as the following sum of rook coefficients.  
 \begin{corollary} Let  $\rookp{n+1}{n} = \rook{1}{n+1}$  and  $\rookp{k}{n}=\rook{k}{n}$ if  $k\leq n.$ Then 
 \( |\sfunspace{\M_n}| = \sum_{k=0}^{n+1} \rookp{k}{n}. \) 
\end{corollary}

We conclude this section with another pleasant correspondence between the endomorphisms in $\sfunspace{\M_n}$ and $\Rook{n}{x}$. 
Let $f: \Lat \to \Lat$ be a function over a lattice $(\Lat,\cleq)$. We say that $f$ is \emph{non-reducing} in $\Lat$ iff it does not map any value to a smaller one; i.e. there is no $e \in \Lat$ such that 
$f(e) \cl e.$  The number of join-endomorphisms that are non-reducing in $\M_n$ is exactly the value of the rook polynomial $\Rook{n}{x}$ for $x=1$. 

\begin{corollary}
\label{rook:th}
$\Rook{n}{1} =  \lvert \{ \ f \in \sfunspace{\M_n} \mid f
\mbox{ is non-reducing in }  \M_n \ \} \rvert$.
\end{corollary} 

Table \ref{sfunc-mn-red:table} illustrates the join-endomorphisms over the lattice $\M_n$ as a union $\bigcup_{i=1}^4\fcal_i$. Corollary~\ref{rook:th} follows from the  observation that the set of non-reducing functions in $\M_n$ is equal to $\fcal_4$ whose size is $\Rook{n}{1}$ as shown in the following proof of Theorem~\ref{size4m-n:lemma}. 

\subsubsection{Proof of Theorem~\ref{size4m-n:lemma}.}

We show that $|\sfunspace{\M_n}|$ can be expressed in terms of Laguerre polynomials:
\( |\sfunspace{\M_n}| = (n+1)^2  +  n! \laguerre{n}{-1}\).

Let $\fcal = \bigcup_{i=1}^4\fcal_i$ where the mutually exclusive $\fcal_i$'s
are defined in Table~\ref{sfunc-mn-red:table}, and $I=\{1,\ldots,n\}$. The proof is divided in two parts: (I) $\fcal = \sfunspace{\M_n}$
and (II) $|\fcal| = (n+1)^2  +  n! \laguerre{n}{-1}$.

\begin{table}[ht]
\centering
\setlength{\tabcolsep}{2pt}
\begin{tabular}{| p{0.48\textwidth} | p{0.48\textwidth} |} 
\hline
\begin{center}
\resizebox{0.3\textwidth}{!}{%
\begin{tikzpicture}[>=stealth]
  \tikzstyle{every node}=[font=\Large]
  \node[shape = circle, draw] (b) at (0,-2) {$\bot$};
  \node[shape = circle, draw] (1) at (-4,0) {$1$};
  \node[shape = circle, draw] (2) at (-2,0) {$2$};
  \node[shape = circle, draw] (3) at (0,0)  {$3$};
  \node[shape = circle, draw] (4) at (2,0)  {$4$};
  \node[shape = circle, draw] (5) at (4,0)  {$5$};
  \node[shape = circle, draw] (t) at (0,2)  {$\top$};

  \draw[gray] (b) to node {} (1);
  \draw[gray] (b) to node {} (2);
  \draw[gray] (b) to node {} (3);
  \draw[gray] (b) to node {} (4);
  \draw[gray] (b) to node {} (5);
  \draw[gray] (1) to node {} (t);
  \draw[gray] (2) to node {} (t);
  \draw[gray] (3) to node {} (t);
  \draw[gray] (4) to node {} (t);
  \draw[gray] (5) to node {} (t);
  
  \draw [blue,->,ultra thick,loop below] (b) to node {} (b);
  \draw [blue,->,ultra thick,bend right] (1) to node {} (b);
  \draw [blue,->,ultra thick,bend right] (2) to node {} (b);
  \draw [blue,->,ultra thick,bend right] (3) to node {} (b);
  \draw [blue,->,ultra thick,bend left]  (4) to node {} (b);
  \draw [blue,->,ultra thick,bend left]  (5) to node {} (b);
  \draw [blue,->,ultra thick,bend left]  (t) to node {} (b);
 \end{tikzpicture}
}%
\end{center}
&
\begin{center}
\resizebox{0.3\textwidth}{!}{%
\begin{tikzpicture}[>=stealth]
  \tikzstyle{every node}=[font=\Large]
  \node[shape = circle, draw] (b) at (0,-2) {$\bot$};
  \node[shape = circle, draw] (1) at (-4,0) {$1$};
  \node[shape = circle, draw] (2) at (-2,0) {$2$};
  \node[shape = circle, draw] (3) at (0,0)  {$3$};
  \node[shape = circle, draw] (4) at (2,0)  {$4$};
  \node[shape = circle, draw] (5) at (4,0)  {$5$};
  \node[shape = circle, draw] (t) at (0,2)  {$\top$};

  \draw[gray] (b) to node {} (1);
  \draw[gray] (b) to node {} (2);
  \draw[gray] (b) to node {} (3);
  \draw[gray] (b) to node {} (4);
  \draw[gray] (b) to node {} (5);
  \draw[gray] (1) to node {} (t);
  \draw[gray] (2) to node {} (t);
  \draw[gray] (3) to node {} (t);
  \draw[gray] (4) to node {} (t);
  \draw[gray] (5) to node {} (t);
  
  \draw [blue,->,ultra thick,loop below] (b) to node {} (b);
  \draw [blue,->,ultra thick,loop left]  (1) to node {} (1);
  \draw [blue,->,ultra thick,bend right] (2) to node {} (1);
  \draw [blue,->,ultra thick,bend right] (3) to node {} (1);
  \draw [blue,->,ultra thick,bend right] (4) to node {} (1);
  \draw [blue,->,ultra thick,bend left]  (5) to node {} (b);
  \draw [blue,->,ultra thick,bend right] (t) to node {} (1);
 \end{tikzpicture}
}%
\end{center}
\\
Let $\fcal_1$ be the family of functions $f$ that for all
$e \in \M_n, f(e) = \bot$.
&
Let $\fcal_2$ be the family of bottom preserving functions
$f$ such that for some $e,e' \in I$: 
(a) $f(\top) = e$, (b) $f(e') = \bot$ or $f(e') = e$, and (c) $f(e'') = e$ for all $e'' \in I \setminus \{e'\}$.
\\
\hline
\begin{center}
\resizebox{0.3\textwidth}{!}{%
\begin{tikzpicture}[>=stealth]
  \tikzstyle{every node}=[font=\Large]
  \node[shape = circle, draw] (b) at (0,-2) {$\bot$};
  \node[shape = circle, draw] (1) at (-4,0) {$1$};
  \node[shape = circle, draw] (2) at (-2,0) {$2$};
  \node[shape = circle, draw] (3) at (0,0)  {$3$};
  \node[shape = circle, draw] (4) at (2,0)  {$4$};
  \node[shape = circle, draw] (5) at (4,0)  {$5$};
  \node[shape = circle, draw] (t) at (0,2)  {$\top$};

  \draw[gray] (b) to node {} (1);
  \draw[gray] (b) to node {} (2);
  \draw[gray] (b) to node {} (3);
  \draw[gray] (b) to node {} (4);
  \draw[gray] (b) to node {} (5);
  \draw[gray] (1) to node {} (t);
  \draw[gray] (2) to node {} (t);
  \draw[gray] (3) to node {} (t);
  \draw[gray] (4) to node {} (t);
  \draw[gray] (5) to node {} (t);
  
  \draw [blue,->,ultra thick,loop below] (b) to node {} (b);
  \draw [blue,->,ultra thick,bend right] (1) to node {} (b);
  \draw [blue,->,ultra thick,bend left]  (2) to node {} (t);
  \draw [blue,->,ultra thick,bend left]  (3) to node {} (t);
  \draw [blue,->,ultra thick,bend right] (4) to node {} (t);
  \draw [blue,->,ultra thick,bend right] (5) to node {} (t);
  \draw [blue,->,ultra thick,loop above] (t) to node {} (t);
 \end{tikzpicture}
}%
\end{center}
&
\begin{center}
\resizebox{0.3\textwidth}{!}{%
\begin{tikzpicture}[>=stealth]
  \tikzstyle{every node}=[font=\Large]
  \node[shape = circle, draw] (b) at (0,-2) {$\bot$};
  \node[shape = circle, draw] (1) at (-4,0) {$1$};
  \node[shape = circle, draw] (2) at (-2,0) {$2$};
  \node[shape = circle, draw] (3) at (0,0)  {$3$};
  \node[shape = circle, draw] (4) at (2,0)  {$4$};
  \node[shape = circle, draw] (5) at (4,0)  {$5$};
  \node[shape = circle, draw] (t) at (0,2)  {$\top$};

  \draw[gray] (b) to node {} (1);
  \draw[gray] (b) to node {} (2);
  \draw[gray] (b) to node {} (3);
  \draw[gray] (b) to node {} (4);
  \draw[gray] (b) to node {} (5);
  \draw[gray] (1) to node {} (t);
  \draw[gray] (2) to node {} (t);
  \draw[gray] (3) to node {} (t);
  \draw[gray] (4) to node {} (t);
  \draw[gray] (5) to node {} (t);
  
  \draw [blue,->,ultra thick,loop below] (b) to node {} (b);
  \draw [blue,->,ultra thick,loop left]  (1) to node {} (1);
  \draw [blue,->,ultra thick,bend left]  (2) to node {} (3);
  \draw [blue,->,ultra thick,bend left]  (3) to node {} (4);
  \draw [blue,->,ultra thick,bend right] (4) to node {} (t);
  \draw [blue,->,ultra thick,bend right] (5) to node {} (t);
  \draw [blue,->,ultra thick,loop above] (t) to node {} (t);
 \end{tikzpicture}
}%
\end{center}
\\
Let $\fcal_3$ be the family of top and bottom preserving functions
$f$ such that for some $e \in I$: (a) $f(e) = \bot$, and (b) $f(e') = \top$ for all $e' \in I \setminus \{e\}$.
&
Let $\fcal_4$ be the family of top and bottom preserving functions 
$f$ that for some $J \subseteq I$:

(a) $f(e) = \top$ for every $e \in J$,
(b) $\fres{f}{I \setminus J}$ is injective, and
(c) $\mathrm{Img}(\fres{f}{I \setminus J})\subseteq I$.
\\
\hline
\end{tabular}
\caption{Families $\fcal_1,\ldots,\fcal_4$ of join-endomorphisms of $\M_n.$ $I = \{1, \dots, n\}.$  $\fres{f}{A}$ is the restriction of $f$ to a subset $A$ of its domain. 
$\mathrm{Img}(f)$ is the image of $f$. A function from each $\fcal_i$ for $\M_5$ is depicted with blue arrows.}
\label{sfunc-mn-red:table}
\end{table}
\subsubsection*{Part (I)}
For $\fcal \subseteq \sfunspace{\M_n}$, it is easy to verify that each $f \in \fcal$ is a join-endomorphism.

For $\sfunspace{\M_n} \subseteq \fcal$ we show that for any function $f$ from $\M_n$ to $\M_n$  if $f \not\in \fcal$, then $f \not\in \sfunspace{\M_n}$. Immediately, if $f(\bot)\neq \bot$ then $f \not\in  \sfunspace{\M_n}$. 

Suppose $f(\bot) = \bot$.
Let $J, K, H$ be disjoint possibly empty sets such that $I = J \cup K \cup H$ and let $j= |J|$, $k = |K|$ and $h = |H|$. The sets $J, K, H$ represent the elements of $I$ mapped by $f$ to $\top$, to elements of $I$, and to $\bot$, respectively. More precisely,
$\mathrm{Img}(\fres{f}{J}) = \{\top\}$,
$\mathrm{Img}(\fres{f}{K}) \subseteq I$ and
$\mathrm{Img}(\fres{f}{H}) = \{\bot\}$.
Furthermore, for every $f$ either (1) $f(\top) = \bot$, (2) $f(\top) \in I$ or (3) $f(\top) = \top$. For each case we show that $f \not\in  \sfunspace{\M_n}$.

\begin{enumerate}
  \item $f(\top) = \bot$. Since $f \not\in \fcal_1$ there is an $e \in I$ such that $f(e) \neq \bot$. We have $e \cleq \top$ but $f(e) \not\cleq f(\top)$. Then $f$ is not monotonic. From Proposition~\ref{prop:monotonicity} we conclude  $f \not\in \sfunspace{\M_n}$.
  
  \item $f(\top) \in I$. Let $K_1,K_2$ be disjoint possibly empty sets such that $K_1 \cup K_2 = K$, $\mathrm{Img}(\fres{f}{K_1}) = \{f(\top)\}$ and $\mathrm{Img}(\fres{f}{K_2}) \neq \{f(\top)\}$. Notice that if $j > 0$ or $|K_2| > 0$, $f$ is non-monotonic and then $f \not\in \sfunspace{\M_n}$.
  
  We then must have $j = 0$ and $K_2 = \emptyset$. Since $\mathrm{Img}(\fres{f}{K}) = \{f(\top)\}$ and $f \not\in \fcal_2$ then $h > 1$. Therefore there must be $e_1, e_2 \in H$ such that $f(e_1) = f(e_2) = \bot$. This implies $f(e_1 \join e_2) = f(\top) \neq \bot = f(e_1) \join f(e_2)$, therefore $f \not\in \sfunspace{\M_n}$.
  
  \item $f(\top) = \top$. 
  
  \begin{enumerate}[label*=\arabic*.]
  \item Suppose $k = 0$. Notice that $f \not\in \fcal_3$ and $f \not\in \fcal_4$ hence $h \neq 1$ and $h \neq 0$. Thus $h >1$ implies that there are at least two $e_1, e_2 \in H$ such that $f(e_1) = f(e_2) = \bot$. But then  $f(e_1 \join e_2) = f(\top) = \top \neq \bot = f(e_1) \join f(e_2)$, hence $f \not\in \sfunspace{\M_n}$.
  
  \item Suppose $k > 0$.
  Assume $h=0$. Notice that $K = I \setminus J$ and $\mathrm{Img}(\fres{f}{K}) \subseteq I$. Since $f$ is a bottom and top preserving function and it satisfies conditions (a) and (c) of $\fcal_4$ but $f \not\in \fcal_4$, then $f$ must violate condition (b). Thus $\fres{f}{K}$ is not injective. Then there are $a,b \in K$ such that $a \neq b$ but $f(a) = f(b)$. Then $f(a) \join f(b) \neq \top = f(a \join b)$. Consequently, $f \not\in \sfunspace{\M_n}$.
  
  \noindent Assume $h>0$. There must be $e_1, e_2, e_3 \in I$  such that $f(e_1) = \bot$ and $f(e_2) = e_3$. Notice that $f(e_1) \join f(e_2) = e_3 \neq \top = f(\top) = f(e_1 \join e_2)$. Therefore, $f \not\in \sfunspace{\M_n}$.
  \end{enumerate}
\end{enumerate}

\subsubsection*{Part (II)}
We prove that $|\fcal| = \sum_{i=1}^4|\fcal_i| = (n+1)^2  +  n! \laguerre{n}{-1}$. Recall that $n = |I|$.

\begin{enumerate}
\item $|\fcal_1| = 1$.
There is only one function mapping every element in $\M_n$ to $\bot$.

\item $|\fcal_2| = n^2 + n$.
There are $n$ possibilities to choose an element of $I$ to map $\top$ to. If there is an element of $I$ mapped to $\bot$, for each one of the previous $n$ options there are also $n$ possibilities to choose such an element. Then, in this case there are $n^2$ functions. If no element of $I$ is mapped to $\bot$, then there are $n$ additional functions.

\item $|\fcal_3| = n$. One of the elements of $I$ is mapped to $\bot$. All the other elements of $I$ are mapped to $\top$. Then, there are $n$ functions in $\fcal_3$.

\item $|\fcal_4| = n!\laguerre{n}{-1} = \Rook{n}{1}$.
Let $f \in \fcal_4$ and let $J \subseteq I$ be a possibly empty set such that $\mathrm{Img}(\fres{f}{J}) = \{\top\}$ and $\mathrm{Img}(\fres{f}{I \setminus J}) \subseteq I$, where $\fres{f}{I \setminus J}$ is an injective function. We shall call $j = |J|$.

For each of the $\binom{n}{j}$ possibilities for $J$, the elements of $I \setminus J$ are to be mapped to $I$ by the injective function $\fres{f}{I \setminus J}$. The number of functions $\fres{f}{I \setminus J}$ is $\frac{n!}{j!}$. Therefore, $|\fcal_4| = \sum_{j=0}^{n} \binom{n}{j} \frac{n!}{j!}$. This sum equals $n! \laguerre{n}{-1}$ which in turn is equal to $\Rook{n}{1}$.
\end{enumerate}

It follows that $|\fcal| = \sum_{i=1}^4|\fcal_i| = (n + 1)^2 + n! \laguerre{n}{-1} $ as wanted.
\qed


\section{Computing the Meet of Join-Endomorphisms}
\label{sec:algo}

We shall provide efficient algorithms for the maximization problem mentioned in the introduction.

\begin{problem}
  \label{prob:meet}
  Given a finite lattice $\Lat$ of size $n$ and  $S \subseteq \sfunspace{L}$ of size $m$, find the \emph{greatest} join-endomorphism $h:\Lat \to \Lat$ below all elements of $S$, i.e. $h = \bigmeetp{\jhspace{\Lat}} S$.
\end{problem}

Notice that the lattice $\jhspace{\Lat}$, which could be exponentially bigger than $\Lat$ (see Section~\ref{sec:size-fs}), is not an input to the problem above.

Finding $\Dfun{S}$ may not be immediate. For instance, see $\Dfun{S}$ in Figure~\ref{fig:difficult-small} for a small lattice of four elements and two join-endomorphisms. As already mentioned, a \emph{naive approach} is to compute $\Dfun{S}$ by taking  $\dapproxfunapp{S}{c} \defsymbol\bigmeetp{\Lat}\{ f(c) \mid f \in S \}$ for each $c \in \Con$. This does not work since $\dapproxfun{S}$ is not necessarily a join-endomorphism as shown in Figure~\ref{fig:meet-pw}.

A \emph{brute force} solution to compute $\Dfun{S}$ can be obtained by generating the set $S' = \{ g \mid g \in \sfunspace{\C} \mbox{ and } g \cleq f \mbox{ for all } f \in S \}$ and taking its join. This approach works since $\bigjoin S' = \Dfun{S}$  but as shown in Section~\ref{sec:size-fs}, the size of $\sfunspace{\C}$ can be super-polynomial for distributive lattices and exponential in general.

Nevertheless, one can use lattice properties to compute
$\Dfun{S}$ efficiently. For distributive lattices, we use the inherent
compositional nature of $\Dfun{S}$. For arbitrary lattices, we present an
algorithm that uses the function $\dapproxfun{S}$ in the naive
approach to compute $\Dfun{S}$ by approximating it
from above.

We will give the time complexities  in
terms of  the number of basic binary lattice operations (i.e. meets, joins
and subtractions) performed during  execution. 

\begin{figure}[t]
\centering
\begin{subfigure}[b]{0.3\textwidth}
\begin{tikzpicture}[scale=0.3,>=stealth]
  \tikzstyle{every node}=[font=\scriptsize]
  \node[shape = circle, draw] (A) at (0,-5)   {$\bot$};
  \node[shape = circle, draw] (B) at (-5,0)   {$1$};
  \node[shape = circle, draw] (C) at (5,0)    {$2$};
  \node[shape = circle, draw] (D) at (0,5)    {$\top$};

  \draw[gray] (A) to node {} (B);
  \draw[gray] (A) to node {} (C);
  \draw[gray] (B) to node {} (D);
  \draw[gray] (C) to node {} (D);

  \draw [blue,->,bend left,dotted,thick]     (B) to node {} (C);
  \draw [blue,->,bend left,dotted,thick]     (C) to node {} (B);
  \draw [blue,->,loop left,dotted,thick]     (A) to node {} (A);
  \draw [blue,->,loop left,dotted,thick]     (D) to node {} (D);
  
  \draw [red,->,loop right,thick]            (D) to node {} (D);
  \draw [red,->,bend left,thick]              (B) to node {} (D);
  \draw [red,->,loop right,thick]            (A) to node {} (A);
  \draw [red,->,loop below,thick]            (C) to node {} (C);

  \draw [teal,dashed,loop below,->,thick]  (A) to node {} (A);
  \draw [teal,dashed,bend left,->,thick]   (D) to node {} (C);
  \draw [teal,dashed,bend left,->,thick]   (C) to node {} (A);
  \draw [teal,dashed,->,thick]             (B) to node {} (C);
\end{tikzpicture}
\caption{{\color{blue} $f : \dottedarrow$}, {\color{red} $g {:} \longrightarrow$}, {\color{teal} $h {:} \dashrightarrow$}}
\label{fig:difficult-small}
\end{subfigure}
~
\begin{subfigure}[b]{0.3\textwidth}
\begin{tikzpicture}[scale=0.3,>=stealth]
  \tikzstyle{every node}=[font=\scriptsize]
  \node[shape = circle, draw] (A) at (0,-5)   {$\bot$};
  \node[shape = circle, draw] (B) at (-5,0)   {$1$};
  \node[shape = circle, draw] (C) at (5,0)    {$2$};
  \node[shape = circle, draw] (D) at (0,5)    {$\top$};

  \draw[gray] (A) to node {} (B);
  \draw[gray] (A) to node {} (C);
  \draw[gray] (B) to node {} (D);
  \draw[gray] (C) to node {} (D);

  \draw [blue,->,bend left,dotted,thick]     (B) to node {} (C);
  \draw [blue,->,bend left,dotted,thick]     (C) to node {} (B);
  \draw [blue,->,loop left,dotted,thick]     (A) to node {} (A);
  \draw [blue,->,loop left,dotted,thick]     (D) to node {} (D);
  
  \draw [red,->,loop right,thick]            (D) to node {} (D);
  \draw [red,->,bend left,thick]             (B) to node {} (D);
  \draw [red,->,loop right,thick]            (A) to node {} (A);
  \draw [red,->,loop below,thick]            (C) to node {} (C);

  \draw [teal,dashed,loop below,->,thick]    (A) to node {} (A);
  \draw [teal,dashed,loop above,->,thick]    (D) to node {} (D);
  \draw [teal,dashed,bend left,->,thick]     (C) to node {} (A);
  \draw [teal,dashed,->,thick]               (B) to node {} (C);
\end{tikzpicture}
\caption{{\color{blue} $f : \dottedarrow$}, {\color{red} $g {:} \longrightarrow$}, {\color{teal} $\dapproxfun{S} {:} \dashrightarrow$}}
\label{fig:meet-pw}
\end{subfigure}
~
\begin{subfigure}[b]{0.3\textwidth}
\begin{tikzpicture}[scale=0.3,>=stealth]
  \tikzstyle{every node}=[font=\scriptsize]
  \node [shape = circle, draw] (A) at (0,-5)   {$\bot$};
  \node [shape = circle, draw] (B) at (-5,0)   {$1$};
  \node [shape = circle, draw] (C) at (0,0)    {$2$};
  \node [shape = circle, draw] (D) at (5,0)    {$3$};
  \node [shape = circle, draw] (E) at (0,5)    {$\top$};

  \draw[gray] (A) to node {} (B);
  \draw[gray] (A) to node {} (C);
  \draw[gray] (A) to node {} (D);
  \draw[gray] (B) to node {} (E);
  \draw[gray] (C) to node {} (E);
  \draw[gray] (D) to node {} (E);

  \draw [blue,loop above,->,dotted,thick]    (B) to node {} (B);
  \draw [blue,bend left,->,dotted,thick]     (C) to node {} (D);
  \draw [blue,bend left,->,dotted,thick]     (D) to node {} (C);
  \draw [blue,loop below,->,dotted,thick]    (A) to node {} (A);
  \draw [blue,loop above,->,dotted,thick]    (E) to node {} (E);
  
  \draw [red,loop right,->,thick]            (E) to node {} (E);
  \draw [red,loop right,->,thick]            (C) to node {} (C);
  \draw [red,bend left,->,thick]             (B) to node {} (E);
  \draw [red,loop below,->,thick]            (D) to node {} (D);
  \draw [red,loop left,->,thick]             (A) to node {} (A);

  \draw [teal,loop right,->,dashed,thick]    (A) to node {} (A);
  \draw [teal,bend right,->,dashed,thick]    (C) to node {} (A);
  \draw [teal,bend left,->,dashed,thick]     (D) to node {} (A);
  \draw [teal,bend right,->,dashed,thick]    (E) to node {} (A);
  \draw [teal,loop below,->,dashed,thick]    (B) to node {} (B);
\end{tikzpicture}
\caption{{\color{blue} $f : \dottedarrow$}, {\color{red} $g {:} \longrightarrow$}, {\color{teal} $\dplusfun{S} {:} \dashrightarrow$}}
\label{ex:m3}
\end{subfigure}
\caption{$S = \{f, g\} \subseteq \jhspace{\Lat}$. (a) $h = \Dfun{S}$. (b) $\dapproxfunapp{S}{c}\defsymbol f(c) \meet g(c)$ is not a join-endomorphism of $\M_2$: $\dapproxfunapp{S}{1 \join 2}\neq \dapproxfunapp{S}{1} \join \dapproxfunapp{S}{2}.$ (c) $\dplusfun{S}$ in Lemma \ref{lemma:delta-ast} is not a join-endomorphism of the non-distributive lattice $\M_3$: $\dplusfunapp{S}{1} \join \dplusfunapp{S}{2} = 1 \neq \bot = \dplusfunapp{S}{1 \join 2}.$ }
\label{fig:mn-lat}
\end{figure}

\subsection{Meet of Join-Endomorphisms in Distributive Lattices}
\label{ssec:dist-lat}

Here we shall illustrate some pleasant compositionality properties of
the infima of join-endomorphisms that can be used for computing the join-endomorphism $\Dfun{S}$ in a finite distributive lattice $\Lat$. In what follows we assume  $n = |\Lat|$ and $m = |S|$. Then we will further improve these results by focusing on the join-irreducible elements of $\Lat$ when computing $\Dfun{S}$.


We use $\tuplespace{X}{J}$ to denote the set of tuples $(x_j)_{j \in J}$ of 
elements $x_j \in X$ for each $j \in J.$ 

\begin{lemma}[\cite{guzman-jlamp-2021}]
\label{lemma:delta-ast}
Let $\Lat$ be a finite distributive lattice and  $S = \{ f_i \}_{i \in I} \subseteq \sfunspace{L}$. Then
\( \Dfun{S}  = \dplusfun{S} \mbox {\ where \ } \dplusfunapp{S}{c} \defsymbol \bigmeetp{\Lat}\{ \bigjoin_{i \in I}\sfunapp{i}{a_i} \mid (a_i)_{i \in I} \in \tuplespace{\Con}{I} \mbox { and }\ \bigjoin_{i \in I}{a_i} \cgeq c \}.\)
\end{lemma}

The above lemma basically says that $\Dfunapp{S}{c}$ is the greatest element in $\Lat$
below all possible applications of the functions in $S$ to elements whose join is greater or equal to $c$.   The proof that $\dplusfun{S} \hgeq \Dfun{S}$ uses the
fact that join-endomorphisms preserve joins. The proof that $\dplusfun{S} \hleq
\Dfun{S}$ proceeds  by showing that  $\dplusfun{S}$  is a lower bound in $\sfunspace{L}$ of $S$. Distributivity of the 
lattice $\Lat$ is crucial for this direction. In fact without it
$\Dfun{S} = \dplusfun{S}$ does not necessarily hold as shown by the following 
counter-example.

\begin{example}
Consider the non-distributive lattice $\mathbf{M}_3$ and $S = \{ f, g \}$ defined  as in Figure~\ref{ex:m3}.
We obtain $\dplusfunapp{S}{1 \join 2} = \dplusfunapp{S}{\top} = \bot$ and
$\dplusfunapp{S}{1} \join \dplusfunapp{S}{2} = 1 \join \bot = 1$. Then,
$\dplusfunapp{S}{1 \join 2} \neq \dplusfunapp{S}{1} \join \dplusfunapp{S}{2}$,
i.e. $\dplusfun{S}$ is not a join-endomorphism.
\end{example}

We rewrite Problem~\ref{prob:meet} to work with only two join-endomorphisms.

\begin{problem}
  \label{prob:meet2}
  Given a lattice $\Lat$ of size $n$ and two join-endomorphisms $f,g:\Lat \to \Lat$, find the \emph{greatest} join-endomorphism $h:\Lat \to \Lat$ below both $f$ and $g$: i.e. $h = f \meetp{\jhspace{\Lat}} g$.
\end{problem}

The following remark gives us compelling reason for using this simplified version of
Problem~\ref{prob:meet}.

\begin{remark}
  \label{rmk:meet-assoc}
  From the associativity of the meet operator $(\meet)$, given a finite lattice $\Lat$ and a set of join-endomorphisms $S = \{ \sfun{i} \mid  i\in I \} \subseteq \jhspace{\Lat}$, where $I = \{1, 2,\ldots, m\}$. We have $\Dfun{S} = (\ldots(\sfun{1} \meetp{\jhspace{\Lat}} \sfun{2}) \meetp{\jhspace{\Lat}} \cdots ) \meetp{\jhspace{\Lat}} \sfun{m}$.
\end{remark}

Thanks to associativity of the meet operator $(\meet)$, solving Problem~\ref{prob:meet} is equivalent to solving Problem~\ref{prob:meet2} for $m$ pairs of join-endomorphisms in $S$.

Nevertheless, we can use Lemma~\ref{lemma:delta-ast} to provide a recursive characterization of $\Dfun{S}$ that can be used in a divide-and-conquer algorithm with lower time complexity.  

\begin{corollary}
  \label{cor:rec1}
  Let $\Lat$ be a finite distributive lattice and  $S = \{f, g\} \subseteq \jhspace{L}$. Then
  \( \Dfun{S}  = \dplusfun{S} \mbox {\ where \ } \dplusfunapp{S}{c} \defsymbol \bigmeetp{\Lat}\{ f(a) \joinp{\Lat} g(b) \mid a,b \in \Lat \mbox { and }\ a \join b \cgeq c \}.\)
\end{corollary}

The above corollary bears witness to the compositional nature of $\Dfun{S}$.

\paragraph{Naive Algorithm $A_1$} One could use Corollary~\ref{cor:rec1} directly in the obvious way, along with Remark~\ref{rmk:meet-assoc} to provide an algorithm for $\Dfun{S}$ by computing $\dplusfun{S}$: i.e. computing, for a pair $f, g \in S,$ the meet of elements of the form $f(a) \joinp{\Lat} g(b)$ for every pair $a,b \in \Lat$ such that  $a \join b \cgeq c.$ And repeating the process for the remaining functions in $S.$ For each $c \in \Lat$,  $\dplusfunapp{\{f,g\}}{c}$ performs $O(n^2)$ meets and $O(n^2)$ joins; and this must be done $m$ times. Thus, $A_1$ can compute $\Dfun{S}$  by performing $O(n \times n^2 \times m) = O(mn^3)$ binary lattice operations.




\subsection{Using Subtraction and Downsets to characterize $\Dfun{S}$}
In what follows we show that  $\Dfun{S}$ can be computed in $O(mn^2)$ for distributive lattices. To achieve this we use the subtraction operator from co-Heyting algebras and the notion of down set\footnote{Recall that we give time complexities  in terms of  the number of basic binary lattice operations (i.e. meets, joins and subtractions) performed during  execution. }.

\paragraph{Subtraction Operator}  Notice that in Corollary~\ref{cor:rec1} we are considering \emph{all} pairs $a,b \in \Lat$ such that $a \join b \cgeq c$. However, because of the monotonicity of join-endomorphisms, it suffices to take, for each $a \in \Lat$, just \emph{the least} $b$ such that $a \join b \cgeq c$. In finite distributive lattices, and more generally in co-Heyting algebras \cite{gierz2003continuous},  the \emph{subtraction} operator $c \latminus a$ gives us exactly such a least element. The subtraction operator  is uniquely determined by the property (\emph{Galois connection}) $b \cgeq c \latminus a$ iff $a \join b \cgeq c$ for all $a,b,c \in \Lat$. 

\paragraph{Down-sets} Besides using just $c \latminus a$ instead of all $b$'s such that $a \join b \cgeq c$, we can use a further simplification: Rather than including every $a \in \Lat,$ we only need to consider every $a$ in the \emph{down-set} of $c$.  Recall that the down-set of $c$ is defined as $\downset{c} = \{ e \in \Lat \mid e \cleq c \}$ (see Definition~\ref{def:covers}).
This additional simplification is justified using properties of distributive lattices to show that for any $a' \in \Lat$, such that $a' \not\cleq c$, there exists $a \cleq c$ such that $f(a) \join  g(c \latminus a) \cleq f(a') \join g(c \latminus a')$.

The above observations lead us to the following theorem. 
  
\begin{theorem}
\label{thm:comp-algo}
Let $\Lat$ be a finite distributive lattice, $S = \{f,g\} \subseteq \sfunspace{L}$ and let $h = \Dfun{S}$. Then \( h(c) = \bigmeetp{\Lat}\{ f(a) \join g(c \latminus a) \mid a \in \downset{c} \}.\)
\end{theorem}

The above result can be used to derive a simple recursive algorithm that, given a finite distributive lattice $\Lat$ and $S\subseteq \sfunspace{L}$,  computes $\Dfun{S}$ in worst-case time complexity $O(mn^2)$ where $m=|S|$ and $n=|\Lat|$. We show this algorithm next.

\begin{algorithm}
  \small
  \caption{\textsc{DMeet}($\Lat$, $S$) finds $\Dfun{S} = h$, where $\Lat$ is a finite distributive lattice, $S \subseteq \jhspace{\Lat},$ and $|S| \geq 1$}\label{alg:dmeet}
  \begin{algorithmic}[1]
    \If{$S = \{f\}$}
      \Return $f$ \Comment $f \in \jhspace{\Lat}$ 
    \EndIf
    \If{$S = \{f, g\}$} \Comment $f,g \in \jhspace{\Lat}$
      \For{each $c \in \Lat$}
        \State $h(c) \gets \bigmeetp{\Lat}\left\{ f(a) \join g(c \latminus a) \mid a \in \downset{c} \right\}$
      \EndFor
      \State \Return $h$
    \Else
      \State partition $S$ into $S_1$ and $S_2$ such that $|S_1| = 2$
      \State\Return \textsc{DMeet}($\Lat$, $\{$\textsc{DMeet}($\Lat$, $S_1$) $\} \cup S_2$) 
    \EndIf
  \end{algorithmic}
\end{algorithm}

\paragraph{Complexity for Distributive Lattices} Assume that $L$ is a finite distributive lattice of size $n$, and that $S$ is a subset of $\jhspace{\Lat}$ of size $m$. With Algorithm~\ref{alg:dmeet}, \textsc{DMeet} is called $m$ times, and for each $c\in \Lat$ the size of $\downset{c}$ is at most of size $n$. Then, we must do $O(n)$ meets and $O(n)$ joins for each $c \in \Lat$. The worst-case time complexity of \textsc{DMeet} is then in $O(mn^2)$.

Nevertheless, we will show how to achieve a better upper bound when considering the join-irreducible elements of the finite distributive lattice.

\subsection{Using Join-Irreducible Elements}
\label{ssec:algo-dist}

As seen from the beginning of Section~\ref{sec:algo}, if we rewrite $\dapproxfunapp{S}{c} \defsymbol \bigmeetp{\Lat}\{ f(c) \mid f \in S \}$ where $S=\{f, g\}$, we obtain following somewhat appealing equation
\begin{equation}
  \label{eq:naive-eq:1}
  \left(f \meetp{\jhspace{\Lat}} g\right)(a) = f(a) \meetp{\Lat} g(a)
\end{equation}
Sadly, it still \emph{does not} hold in general, as illustrated in the lattices $\M_2$ and $\M_3$ in Figure~\ref{fig:meet-pw} and Figure~\ref{fig:jimeet-nodist}. 

However, it turns out that we can partly use Equation~\ref{eq:naive-eq:1} to obtain a better upper bound. The following lemma states that Equation~\ref{eq:naive-eq:1} holds if $\Lat$ is distributive and $a \in \ijoin{\Lat}$.

\begin{lemma}
\label{lem:jimeet}
Let $\Lat$ be a finite distributive lattice and $f,g \in \jhspace{\Lat}$. Then the following equation holds: 
$\left( f \meetp{\jhspace{\Lat}} g \right)(a) = f(a) \meetp{\Lat} g(a) \mbox{ for every } a \in \ijoin{\Lat}.$
\end{lemma}

\begin{proof}
From Theorem~\ref{thm:comp-algo},
\(({f \meetp{\jhspace{\Lat}} g})(a) =
\bigmeet \left\{ f(a') \join g(a \sop a')\ |\ a' \in \downset{a} \right\}.\)
Note that since  $a \in \ijoin{\Lat}$ if $a' \in \downset{a}$ then $a \sop a' = a$ when $a \neq a'$, and $a \sop a' = \bot$
when $a = a'$. Then,
\begin{align*}
  \{ f(a') \join g(a \sop a')\mid a' \in \downset{a}\} & = \{ f(a') \join g(a \sop a')\mid a' \cl a\} \cup \{f(a) \join g(\bot)\} \\
  & = \{ f(a') \join g(a)\mid a' \cl a\} \cup \{f(a)\} \\
  & = \{ f(a') \join g(a)\mid \bot \cl a' \cl a\} \cup \{f(a), g(a)\}.
\end{align*}
By absorption, we know that $(f(a') \join g(a)) \meet g(a) = g(a)$. Finally, using properties of $\meet$, 
\begin{align*}
  (f \meetp{\jhspace{\Lat}} g)(a)
  & = \bigmeet \left(\{ f(a') \join g(a)\mid \bot \cl a' \cl a\} \cup \{f(a), g(a)\}\right) \\
  & = \bigmeet \{ f(a') \join g(a)\mid \bot \cl a' \cl a\} \meet f(a) \meet g(a) \\
  & = f(a) \meet g(a).
\end{align*}
\end{proof}

\begin{figure}
  \centering
  \begin{tikzpicture}[scale=0.3,>=stealth]
    \tikzstyle{every node}=[font=\scriptsize]
    \node [shape = circle, draw] (A) at (0,-5)   {$\bot$};
    \node [shape = circle, draw] (B) at (-5,0)   {$1$};
    \node [shape = circle, draw] (C) at (0,0)    {$2$};
    \node [shape = circle, draw] (D) at (5,0)    {$3$};
    \node [shape = circle, draw] (E) at (0,5)    {$\top$};
  
    \draw[gray] (A) to node {} (B);
    \draw[gray] (A) to node {} (C);
    \draw[gray] (A) to node {} (D);
    \draw[gray] (B) to node {} (E);
    \draw[gray] (C) to node {} (E);
    \draw[gray] (D) to node {} (E);
  
    \draw [left,loop left,blue,->,dotted,thick]      (B) to node {} (B);
    \draw [above,blue,->,bend left,dotted,thick]     (C) to node {} (D);
    \draw [below,blue,->,bend left,dotted,thick]     (D) to node {} (C);
    \draw [below,loop below,blue,->,dotted,thick]    (A) to node {} (A);
    \draw [above,loop above,blue,->,dotted,thick]    (E) to node {} (E);
    \draw [above,loop right,red,->,thick]            (E) to node {} (E);
    \draw [left,loop left,red,->,thick]              (C) to node {} (C);
    \draw [above,red,->,bend left,thick]             (B) to node {} (E);
    \draw [right,loop right,red,->,thick]            (D) to node {} (D);
    \draw [below,loop left,red,->,thick]             (A) to node {} (A);
  
    \draw [above,loop right,teal,->,dashed,thick]  (A) to node {} (A);
    \draw [left,teal,->,bend right,dashed,thick]   (C) to node {} (A);
    \draw [right,teal,->,bend left,dashed,thick]   (D) to node {} (A);
    \draw [below,loop below,teal,->,dashed,thick]  (B) to node {} (B);
  \end{tikzpicture}
  \caption{{\color{blue} $f {:} \dottedarrow$}, {\color{red} $g {:} \rightarrow$}, {\color{teal} $h {:} \dashrightarrow$}. Any $h:\M_3 \to \M_3$ s.t. $h(a) = f(a) \meet g(a)$ for $a \in \ijoin{\M_3}$ is not in $\jhspace{\M_3}$: $h(\top) = h(1 \join 2) = h(1) \join h(2) = 1 \neq \bot = h(2) \join h(3) = h(2 \join 3) = h(\top)$.}
  \label{fig:jimeet-nodist}
\end{figure}

It is worth noting the Lemma~\ref{lem:jimeet} may not hold for non-distributive lattices. This is illustrated in Figure~\ref{fig:jimeet-nodist} with 
the archetypal non-distributive lattice $\M_3$. Suppose that $f$ and $g$ are given as in Figure~\ref{fig:jimeet-nodist}.
Let $h =   f \meetp{\jhspace{\Lat}} g$ with  $h(a) = f(a) \meet g(a)$ for all $a \in \{1,2,3\} =\ijoin{\M_3}$. Since $h$ is a join-endomorphism, we would have $h(\top) = h(1 \join 2) = h(1) \join h(2) = 1 \neq \bot = h(2) \join h(3) = h(2 \join 3) = h(\top)$, a contradiction. 

The following proposition will come in handy for defining join-endomorphisms in terms of join-irreducible elements. Recall that $\jdown{e} = \downset{e} \cap \ijoin{\Lat}$ (see Definition~\ref{def:covers}).

\begin{proposition}[\cite{gratzer-latjoinend-1958,davey2002introduction}]
\label{prop:jendo-proprts6}
Let $\Lat$ be a lattice. If $\Lat$ is finite and distributive, $f \in \jhspace{\Lat}$ iff $(\forall e \in \Lat)\ f(e) = \bigjoin \{ f(e')\ |\ e' \in \jdown{e}\}$.
\end{proposition}

Lemma~\ref{lem:jimeet} and~\ref{prop:jendo-proprts6} lead us to the following characterization of meets over $\jhspace{\Lat}.$

\begin{theorem}
  \label{thm:meet-ch}
  Let $\Lat$ be a finite distributive lattice and $f,g \in \jhspace{\Lat}$.   Then $h = f \meetp{\jhspace{\Lat}} g$ iff $h$ satisfies 
  \begin{equation}
  \label{eq:h}
  h(a) =
  \begin{cases}
  f(a) \meet_\Lat g(a) 		\quad & 	  \mbox{ if } a \in \ijoin{\Lat}  \mbox{ or } a = \bot \\
  h(b)  \join_\Lat  h(c)           	\quad &       \mbox{ if } b,c \in \covers{a} \mbox{ with } b\neq c\\
  \end{cases}
  \end{equation}
\end{theorem}

\begin{proof}
The only-if direction follows from Lemma~\ref{lem:jimeet} and Proposition~\ref{prop:jendo-proprts6}.  For the if-direction, suppose that $h$ satisfies Equation~\ref{eq:h}. If $h \in \jhspace{\Lat}$ the result follows from Lemma~\ref{lem:jimeet} and~\ref{prop:jendo-proprts6}. To prove $h \in \jhspace{\Lat}$ from~\ref{prop:jendo-proprts6} it suffices to show 
\begin{equation}
\label{eq:prop-ind-h}
  h(e) = \bigjoin \{ h(e') \mid e' \in \jdown{e}\}
\end{equation}
for every $e \in \Lat$.  From Equation~\ref{eq:h} and since $f$ and $g$ are monotonic, $h$ is monotonic. If $e \in \ijoin{\Lat}$ then $h(e') \cleq h(e)$ for every $e' \in \jdown{e}$. Therefore, $\bigjoin \{ h(e') \mid e' \in \jdown{e}\} = h(e)$.
If $e \not\in \ijoin{\Lat}$, we proceed by induction.  Assume Equation~\ref{eq:prop-ind-h} holds for all $a \in \covers{e}$. By definition, $h(e) = h(b) \join h(c)$ for any $b,c \in \covers{e}$ with $b \neq c$. Then, we have $h(b) = \bigjoin \{ h(e') \mid e' \in \jdown{b}\}$ and $h(c) = \bigjoin \{ h(e') \mid e' \in \jdown{c}\}$. Notice that $e' \in \jdown{b}$ or $e' \in \jdown{c}$ iff $e' \in \jdown{(b \join c)}$, since $\Lat$ is distributive.
Thus, $h(e) = h(b) \join h(c) = \bigjoin \{ h(e') \mid  e' \in \jdown{(b \join c)}\} = \bigjoin \{ h(e') \mid  e' \in \jdown{e}\}$ as wanted.
\end{proof}

We conclude this section by stating the time complexity $O(n)$ to compute $h$ in the above theorem. The time complexity is 
determined by the number of basic binary lattice operations (i.e. meets and joins) performed during execution. 

\begin{corollary}  
\label{thm:jecomplexity}
Given a distributive lattice $\Lat$ of size $n$, and functions
$f, g \in \jhspace{L}$, the function $h = f \meetp{\jhspace{\Lat}} g$ can be 
computed in $O(n)$ binary lattice operations.
\end{corollary}

\begin{proof}
  If $a \in \ijoin{L}$ then from 
Theorem~\ref{thm:meet-ch}, $h(a)$ can be computed as $f(a) \meet g(a)$. If $a =\bot$ then $h(a)$ is $\bot$. 
If $a \notin \ijoin{\Lat}$ and $a \neq \bot$, we pick any $b, c \in \covers{a}$ such that $b \neq c$
and compute $h(a)$ recursively as $h(b) \join h(c)$ by Theorem~\ref{thm:meet-ch}.
We can use a lookup table to keep track of the values of $a \in \Lat$ for which $h(a)$ has been computed, starting with all $a \in \ijoin{\Lat}$.
Since $h(a)$ is only computed once for each $a \in \Lat$, either as a meet for elements in $\ijoin{\Lat}$ or as a join otherwise,
we only perform $n$ binary lattice operations.
\end{proof}

Following Remark~\ref{rmk:meet-assoc} and Corollary~\ref{thm:jecomplexity}, for a finite distributive lattice $\Lat$ of size $n$ and a set of join-endomorphisms $S \subseteq \jhspace{\Lat}$ of size $m$, we can compute $\Dfun{S}$ with a worst-case complexity in $O(mn)$ binary lattice operations.


\subsubsection*{Profiling and Runtime}
\label{ssec:exp-dist}

For distributive lattices, we now present some experimental results comparing the average runtime between Algorithm~\ref{alg:dmeet}  based on Theorem~\ref{thm:comp-algo}, referred to as \textsc{DMeet}, and the proposed algorithm in Theorem~\ref{thm:meet-ch}, called \textsc{DMeet+}.

We use the algorithms presented above to compute the greatest dilation below a given set of dilations and illustrate its result for a simple image in Section~\ref{ssec:ex-mm}.

\begin{figure}
    \centering
    \begin{subfigure}[b]{0.45\textwidth}
        \includegraphics[width=\textwidth]{\mainDir/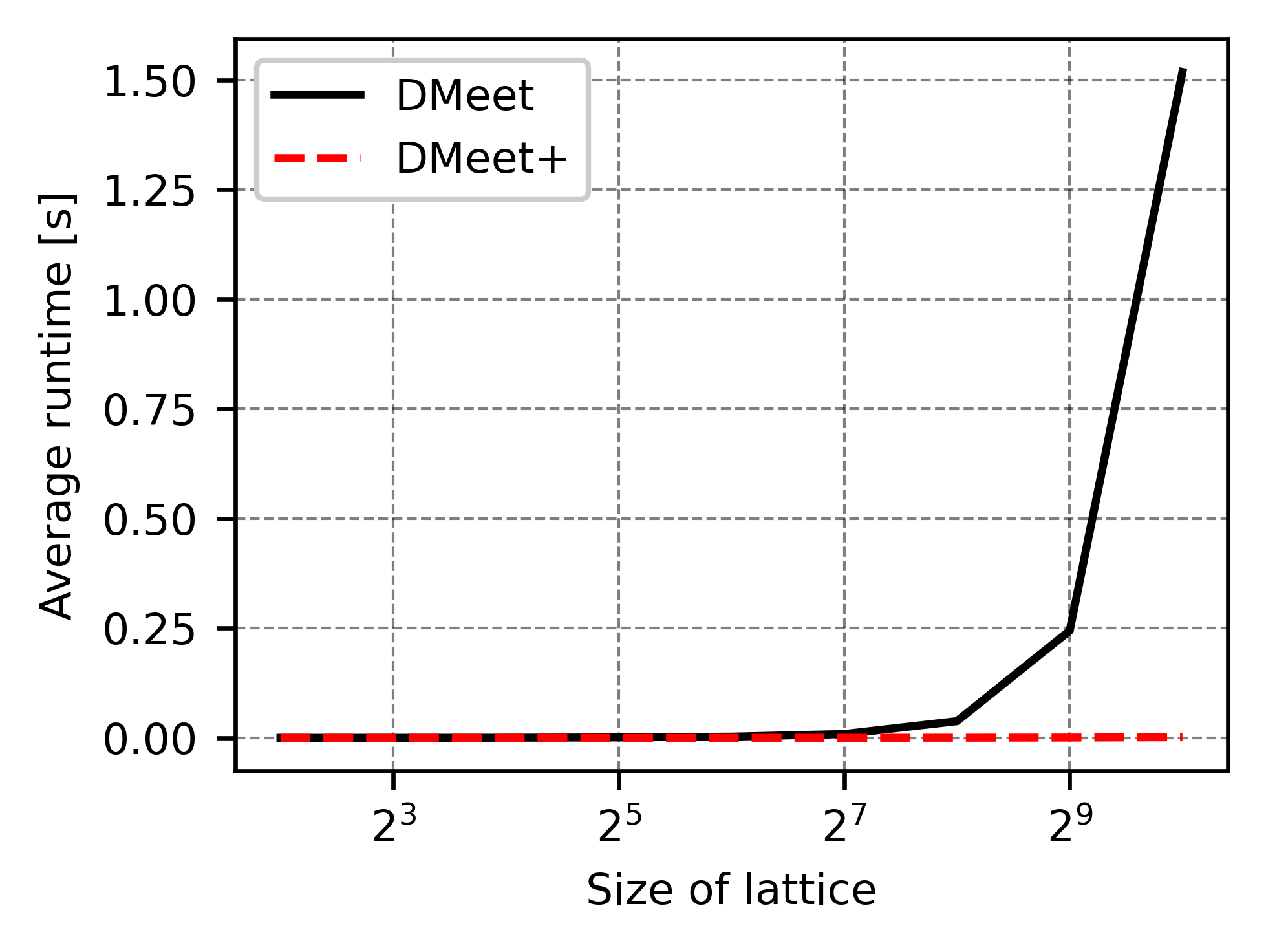}
        \caption{Powerset lattices.}
        \label{fig:dmeet-powerset}
    \end{subfigure}
    ~
    \begin{subfigure}[b]{0.45\textwidth}
        \includegraphics[width=\textwidth]{\mainDir/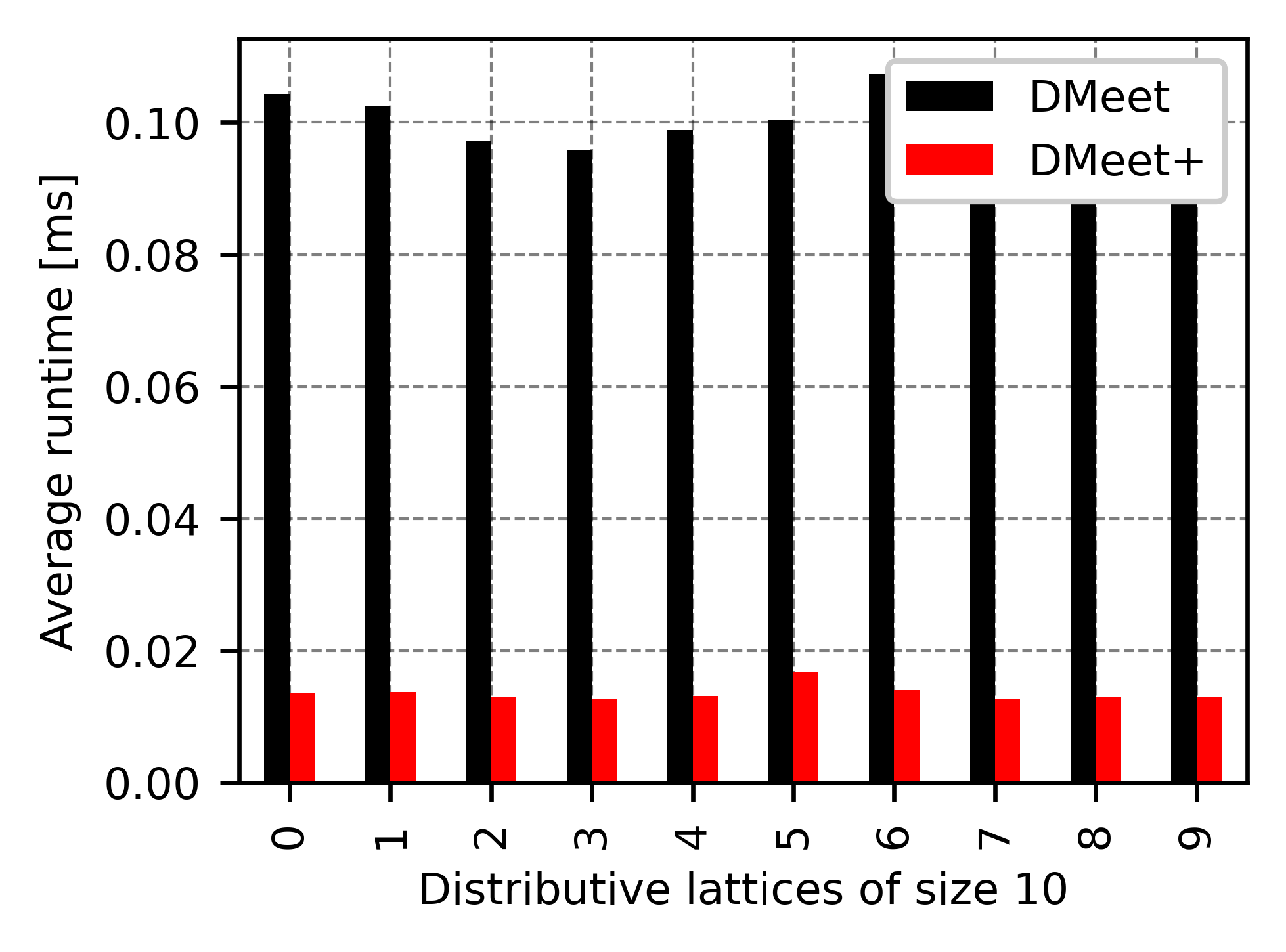}
        \caption{Arbitrary distributive lattices.}
        \label{fig:dmeet-dist}
    \end{subfigure}
\caption{Comparison between an implementation of Theorem~\ref{thm:comp-algo} (\textsc{DMeet})
and Theorem~\ref{thm:meet-ch} (\textsc{DMeet+}).}
\label{fig:dmeet-experiments}
\end{figure}

The algorithms were implemented using Python 3, the code is available in \cite{website:github-delta}. All the results in Figure~\ref{fig:dmeet-experiments}, were executed in a MacBook Pro (Retina, 15-inch, Mid 2014) with a Quad-Core Intel Core i7 processor, and 16 GB of 1600 MHz DDR3 RAM with a constant supply of electricity.

Figure~\ref{fig:dmeet-experiments} shows the average runtime of each algorithm, from 100 runs with a random pair of join-endomorphisms. For Figure~\ref{fig:dmeet-powerset}, we compared each algorithm against powerset lattices of sizes between $2^2$ and $2^{10}.$
For Figure~\ref{fig:dmeet-dist}, 10 random distributive lattices of size 10 were selected. In both cases, all binary lattice
operation are guaranteed a complexity in $O(1)$ to showcase the quadratic nature of \textsc{DMeet} compared to the linear growth of \textsc{DMeet+}.
The time reduction from \textsc{DMeet} to \textsc{DMeet+} is also reflected in a reduction on the number of $\join$ and $\meet$ operations performed as illustrated in Table~\ref{tab:dmeet-experiments}.
For  \textsc{DMeet+}, given a distributive lattice $\Lat$ of size $n,$  \#$\meet = |\ijoin{\Lat}|$
and \#$\join = |\Lat| - |\ijoin{\Lat}| - 1$ ($\bot$ is directly mapped to $\bot$). 

\begin{table}
\begin{center}
\setlength{\tabcolsep}{5pt}
\begin{tabular}{ l r r r r r r }
    \toprule
    {} & \textsc{DMeet} & \textsc{DMeet+} & \textsc{DMeet} & \textsc{DMeet+} & \textsc{DMeet} & \textsc{DMeet+} \\
    Size &  Time [s] &  Time [s] &  \#$\join$ &  \#$\join$  &  \#$\meet$  &  \#$\meet$ \\
    \midrule
    16    &    0.000246 &   0.000024 &          81 &         11 &          81 &          4 \\
    32    &    0.000971 &   0.000059 &         243 &         26 &         243 &          5 \\
    64    &    0.002659 &   0.000094 &         729 &         57 &         729 &          6 \\
    128   &    0.008735 &   0.000163 &        2187 &        120 &        2187 &          7 \\
    256   &    0.038086 &   0.000302 &        6561 &        247 &        6561 &          8 \\
    512   &    0.244304 &   0.000645 &       19683 &        502 &       19683 &          9 \\
    1024  &    1.518173 &   0.001468 &       59049 &       1013 &       59049 &         10 \\
    \bottomrule
\end{tabular}
\caption{Average runtime in seconds over powerset lattices. Number of $\join$ and $\meet$ operations performed for each algorithm.}
\label{tab:dmeet-experiments}
\end{center}
\end{table}

\subsection{Algorithms for Arbitrary Lattices}
\label{ssec:algo-gnrl}

The previous algorithms may fail to produce $\Dfun{S}$ for non-distributive finite lattices.
Nonetheless, for any arbitrary finite lattice $\Lat$, $\Dfun{S}$ can be
computed by successive approximations, starting with some self-map known to be smaller than each  $\sfun{}\in S$ and greater than $\Dfun{S}$.
Assume a self-map $\dapprox: \Con \rightarrow \Con$ such that 
$\dapprox \hgeq\Dfun{S}$,
and for all $\sfun{}\in S$, $\dapprox \hleq \sfun{}$. A good starting
point is $\dapprox(u)=\bigmeet \{\sfun{}(u) \mid \sfun{}\in S\}$, for all
$u\in L$. By definition of meet operator, $\dapprox$ is the biggest function under
all functions in $S$, hence $\dapprox \hgeq \Dfun{S}$.
The program \textsc{GMeet} in Algorithm~\ref{alg:quicksp} computes decreasing upper bounds of $\Dfun{S}$ by correcting
$\dapprox$ values not conforming to the \emph{join-endomorphism property:}  \( \dapprox(u)
\join\dapprox(v)=\dapprox(u\join v).\) The correction decreases $\dapprox$ and
maintains the invariant  $\dapprox \hgeq \Dfun{S}$, as stated in Theorem~\ref{theorem:approx-correct}. The proof of this statement is given in Section~\ref{sec:proof-camilo-thm}.
\begin{theorem}
\label{theorem:approx-correct}
Let  $L$ be a finite  lattice, $u,v\in \Lat$, $\dapprox: \Con \rightarrow \Con$ and  $S \subseteq \sfunspace{L}$. Assume  $\dapprox \hgeq \Dfun{S}$ holds, and consider the following updates:
\begin{enumerate}
\item  when $\dapprox(u)
\join\dapprox(v)\cl\dapprox(u\join v)$, assign
$\dapprox(u\join v) \leftarrow \dapprox(u)\join\dapprox(v)$
\item when 
$\dapprox(u)\join\dapprox(v)\not\cleq\dapprox(u\join v)$, assign
$\dapprox(u)\gets \dapprox(u)\meet\dapprox(u\join v)$ and also 
$\dapprox(v)\gets \dapprox(v)\meet\dapprox(u\join v)$
\end{enumerate}
Let $\dapprox'$ be the function resulting after the update. Then, (1) $\dapprox' \cl \dapprox$ and (2) $\dapprox' \hgeq \Dfun{S}$.
\end{theorem}

\begin{algorithm}
  \small
  \caption{\textsc{GMeet} finds $\dapprox = \Dfun{S}$}\label{alg:quicksp}
  \begin{algorithmic}[1]
    \State $\dapprox(u)\gets\bigmeet \{ f(u) ~|~ f \in S\}$ \Comment{for all $u\in \Con$}
    \While{$u,v\in \Con\wedge \dapprox(u)\join \dapprox(v) \neq \dapprox(u\join v) $}
      \If{$  \dapprox(u)\join\dapprox(v)\cl \dapprox(u\join v)$}\Comment{case (1)}
        \State $\dapprox(u\join v)\gets \dapprox(u)\join \dapprox(v)$
      \Else\Comment{case (2)}
        \State $\dapprox(u)\gets \dapprox(u)\meet\dapprox(u\join v)$
        \State $\dapprox(v)\gets \dapprox(v)\meet\dapprox(u\join v)$
      \EndIf
    \EndWhile
  \end{algorithmic}
\end{algorithm}
The procedure (see Algorithm~\ref{alg:quicksp}) loops through pairs
$u,v\in\Con$ while there is some pair satisfying cases (1) or (2) above for
the current $\dapprox$. When there is, it updates $\dapprox$ as mentioned in Theorem~\ref{theorem:approx-correct}. At the end of the loop all pairs $u,v\in\Con$ satisfy the join preservation
property. By the invariant mentioned in the theorem, this means $\dapprox=
\Dfun{S}$.

As for the previous algorithms in this paper, the worst-time time complexity will be expressed in terms of the binary lattice operations performed during execution. 
Assume a fixed set $S$ of size $m$. The complexity of the initialization (Line 1) of  $\textsc{GMeet}$ is $O(nm)$ with
$n=|\Con\, |$. The value of $\dapprox$ for a given $w\in \Lat$ can be updated (decreased) at most $n$ times. Thus, there are at most $n^2$ updates of $\dapprox$ for all values of $\Lat$. Finding a $w= u \join v$ where $\dapprox(w)$ needs an update because  $\dapprox(u)\join \dapprox(v) \neq \dapprox(u\join v)$ (test of the loop, Line 2) takes $O(n^2)$. Hence, the worst time complexity
of the loop is in $O(n^4)$.

The program \textsc{GMeet+}, in Algorithm~\ref{alg:quicksp2}, uses appropriate data structures to reduce significantly the time complexity of the algorithm. Essentially, different sets are used to keep track of properties of $(u,v)$ lattice pairs with respect to the current $\dapprox$. We have a support (correct) pairs set $\Sseta{w}=\{(u,v)~|~w=u\sqcup v\wedge \dapprox(u)\join \dapprox(v)=\dapprox(w)\}$. We also have a conflicts set $\Cseta{w}=\{(u,v)~|~w=u\sqcup v\wedge \dapprox(u)\join \dapprox(v)\cl\dapprox(w)\}$ and failures set $\Fseta{w}=\{(u,v)~|~w=u\join v\wedge \dapprox(u)\join \dapprox(v)\not\cleq\dapprox(w)\}.$
\begin{algorithm}[t]
\small
\caption{\textsc{GMeet+} finds $\dapprox = \Dfun{S}$}\label{alg:quicksp2}
\begin{algorithmic}[1]
  \State $\dapprox(u)\gets\bigmeet \{ f(u) ~|~ f \in S\}$ \Comment{for all $u\in \Con$}
  \State Initialize $\Sseta{w}, \Cseta{w}, \Fseta{w}$, for all $w$
  \While{$w\in L$ such that $(u,v)\in \Cseta{w}$} \Comment{some conflict set not empty}
    \State $\Cseta{w} \gets \Cseta{w}\backslash \{(u,v)\}$
    \State $\dapprox(w)\gets \dapprox(u)\join \dapprox(v)$ 
    \State $\Fseta{w} \gets \Fseta{w}\cup \Sseta{w}$ \Comment{all pairs previously in $\Sseta{w}$ are now failures}
    \State $\Sseta{w} \gets \{(u,v)\}$ 
    \State \textsc{checkSupports}($w$)\Comment{for $u\in L$, verify property  $\Sseta{w\join u}$}
    \While {$z\in L$ such that $(x,y)\in \Fseta{z}$}  \Comment{some failures set not empty}
      \State $\Fseta{z}\gets \Fseta{z}\backslash\{(x,y)\}$
      \If{$\dapprox(x) \neq  \dapprox(x)\meet\dapprox(z)$}
        \State $\dapprox(x)\gets \dapprox(x)\meet\dapprox(z)$  \Comment{$\dapprox(x)$ decreases}
        \State $\Fseta{x}\gets \Fseta{x}\cup \Sseta{x}$  \Comment{all pairs in $\Sseta{x}$ are now failures}
        \State $\Sseta{x}\gets \emptyset$
        \State \textsc{checkSupports}($x$)\Comment{for $u\in L$, verify property  $\Sseta{x\join u}$}
      \EndIf
      \If{$\dapprox(y) \neq  \dapprox(y)\meet\dapprox(z)$}
        \State $\dapprox(y)\gets \dapprox(y)\meet\dapprox(z)$ \Comment{$\dapprox(y)$ decreases}
        \State $\Fseta{y}\gets \Fseta{y}\cup \Sseta{y}$ \Comment{all pairs in $\Sseta{y}$ are now failures}
        \State $\Sseta{y}\gets \emptyset$
        \State \textsc{checkSupports}($y$)\Comment{for $u\in L$, verify property  $\Sseta{y\join u}$}
      \EndIf
      \If{$\dapprox(x)\join\dapprox(y)=\dapprox(z) $}
        \State $\Sseta{z}\gets \Sseta{z}\cup \{ (x,y)\}$ \Comment{$(x,y)$ is now correct}
      \Else
        \State $\Cseta{z}\gets \Cseta{z}\cup \{(x,y)\}$ \Comment{$(x,y)$ is now a conflict}
      \EndIf
    \EndWhile
  \EndWhile
\end{algorithmic}
\end{algorithm}
Algorithm~\ref{alg:quicksp2} updates $\dapprox$ as mentioned in Theorem~\ref{theorem:approx-correct} and so maintains the invariant $\dapprox \cgeq \Dfun{S}$. An additional invariant is that, for all $w$,  sets $\Sseta{w},\Cseta{w},\Fseta{w}$ are pairwise disjoint. When the outer loop finishes sets $\Cseta{w}$ and $\Fseta{w}$ are empty (for all $w$) and thus every $(u,v)$ belongs to $\Sseta{u\join v}$, i.e. the resulting $\dapprox=\Dfun{S}$.

Auxiliary procedure $\textsc{checkSupports}(u)$ identifies all the pairs $(u,x) \in \Sseta{u \join x}$  that may no longer satisfy the join-endomorphism property \( \dapprox(u) \join \dapprox(x) = \dapprox(u \join x)\) because of an update to $\dapprox(u)$. When this happens, it adds $(u,x)$ to the appropriate $\Cset$, or $\Fset$ set.  The time complexity of the algorithm depends on the set operations computed for each $w\in L$ chosen, either in the \emph{conflicts} $\Cseta{w}$ set or in the \emph{failures} $\Fseta{w}$ set. When a $w$ is selected (for some $(u,v)$ such that $u \join v = w$) the following holds: (1) at least one of $\dapprox(w), \dapprox(u), \dapprox(v)$ is decreased, (2)  some fix $k$ number of elements are removed from or added to a set, (3) a union of two \emph{disjoint} sets is computed, and (4) new support sets of $w, u$ or $v$ are calculated.

\paragraph{Complexity for Arbitrary Lattices}
With an appropriate implementation, operations (1)-(2) take $O(1)$, and also operation (3), since sets are disjoint. Operation (4) clearly  takes $O(n)$. In each loop of the (outer or inner) cycles of the algorithm, at least one reduction of $\dapprox$ is computed. Furthermore, for each reduction of $\dapprox$, $O(n)$ operations are performed. The maximum possible number of $\dapprox(w)$ reductions, for a given $w$, is equal to the length $d$ of the longest strictly decreasing chain in the lattice. The total number of possible $\dapprox$ reductions is thus equal to $nd$. The total number of operations of the algorithm is then $O(n^2 d)$. In general, $d$ could be (at most) equal to $n$, therefore, after  initialization, worst case complexity is $O(n^3)$. The initialization (Lines 1-2) takes $O(nm)+O(n^2)$, where $m = | S |$. Worst time complexity is thus $O(mn + n^3)$. For powerset lattices, $d = \log_2 n$, thus worst time complexity in this case is $O(mn+n^2 \log_2 n)$.

\subsubsection*{Experimental Results}
\label{sec:exp-gnrl}

Here we present some experimental results showing the execution time of the proposed algorithms.

\begin{figure}[t]
\centering
\includegraphics[scale=0.5]{\mainDir/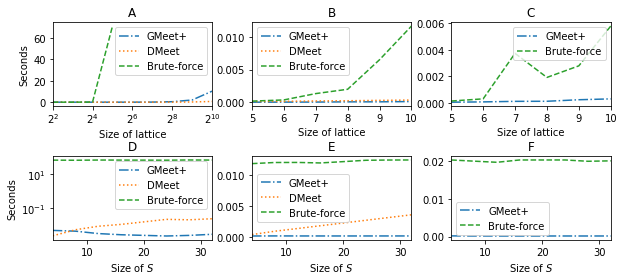}
\caption{\footnotesize {Average performance time of \textsc{GMeet+}, \textsc{DMeet} and \textsc{Brute-force}. Plots A and D use $\mathbf{2}^n$ lattices, B and E distributive lattices, and C and F arbitrary lattices. Plots A-C have a fixed number of join-endomorphisms and plots D-F have a fixed lattice size.}}
\label{fig:exp1}
\end{figure}

Consider Figure~\ref{fig:exp1}. In plots \ref{fig:exp1}.A-C,
the horizontal axis is the size of the lattice.
In plots \ref{fig:exp1}.D-F, the horizontal axis is the size of $S \subseteq \sfunspace{\Lat}$.
Curves in images \ref{fig:exp1}.A-C plot, for each algorithm, the average execution time of 100 runs (10 for \ref{fig:exp1}.A) with 
random sets $S$ of size 4.
Images \ref{fig:exp1}.D-F, show the mean execution time of each algorithm for
100 runs (10 for \ref{fig:exp1}.D) varying the number of join-endomorphisms 
($|S|=4i$, $1\leq i \leq 8$). The lattice size is fixed: $|\Lat| = 10$ for \ref{fig:exp1}.E and \ref{fig:exp1}.F,  and
$|\Lat| = 2^{5} $ for \ref{fig:exp1}.D. In all cases the lattices were randomly generated,
and the parameters selected to showcase the difference between each algorithm with a sensible overall execution time.
For a given lattice $\Lat$ and $S \subseteq \sfunspace{\Lat}$, 
the brute-force algorithm explores the whole space $\sfunspace{\Lat}$ to find all the join-endomorphism 
below each element of $S$ and then computes the greatest of them. In particular, the measured 
spike in plot~\ref{fig:exp1}.C corresponds to the random lattice of seven elements with the size of $\sfunspace{\Lat}$ being  bigger 
than in the other experiments in the same figure. 
In our experiments we observed that for a fixed $S$, as the size of the lattice increases, \textsc{DMeet} outperforms \textsc{GMeet+}. This is noticeable in lattices $\mathbf{2}^n$ (see \ref{fig:exp1}.A).  Similarly, for a fixed lattice, as the size 
of  $S$ increases  \textsc{GMeet+}  outperforms \textsc{DMeet}. 
\textsc{GMeet+} performance can actually improve with a higher number of 
join-endomorphisms (see \ref{fig:exp1}.D) since  the initial $\dapprox$ is usually smaller in this case.

To illustrate some performance gains, Table~\ref{tab:exp1} shows the mean execution 
time of the algorithms discussed in this paper. We include naive algorithm $A_1$ outlined just after Corollary~\ref{cor:rec1}.

\begin{table}[t]
    \begin{center}
    \footnotesize
    \resizebox{0.8\textwidth}{!}{%
        \setlength{\tabcolsep}{15pt}
        \begin{tabular}{ r  r  r  r  r }
            \hline
            Size & $A_1$ & \textsc{GMeet} & \textsc{GMeet}+ & \textsc{DMeet} \\
            \hline
            16 & 2.01 & 0.00360 & 0.000603 & 0.000632 \\
            32 & 64.6 & 0.0633 & 0.00343 & 0.00181 \\
            64 & 1901 & 0.948 & 0.0154 & 0.00542 \\
            128 & \textgreater600 & 15.4 & 0.0860 & 0.0160 \\
            256 & \textgreater600 & 252 & 0.361  & 0.0483\\
            512 & \textgreater600 & \textgreater600 & 2.01 & 0.166 \\
            1024 & \textgreater600 & \textgreater600 & 10.7 & 0.547 \\
            \hline
        \end{tabular}
        }%
        \caption{Average time in seconds over powerset lattices with $|S| = 4$ } 
        \label{tab:exp1}
    \end{center}    
\end{table}


\subsection{Meet of Join-Endomorphisms in Modular lattices}
\label{sec:mod-lat}

In the latter section we have shown that given a finite lattice $\Lat$ and $S \subseteq \jhspace{\Lat}$, Algorithm~\ref{alg:quicksp2} finds $\Dfun{S}$ by starting at $\dapprox(c) = \bigmeet \{\sfun{}(c) \mid \sfun{}\in S\}$ (for all $c \in \Lat$) and correcting the values that violates the equality $\dapprox(a \join b) = \dapprox(a) \join \dapprox(b)$, for every pair $(a,b) \in \Lat^2$. However, one may wonder whether the algorithm tries to correct a value of the function that has been already corrected. Indeed, this may happen for pairs of elements whose join is the same. The following example illustrates such a situation.

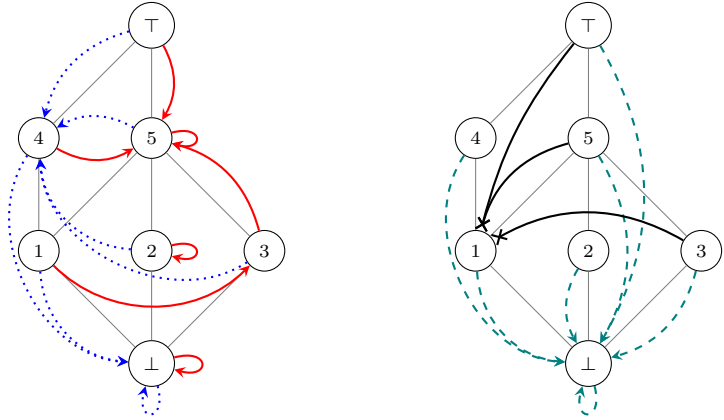
\begin{figure}[t]
  \centering
  \begin{subfigure}[b]{0.45\textwidth}
  \centering
  \begin{tikzpicture}[scale=0.3,>=stealth]
    \tikzstyle{every node}=[font=\scriptsize]
    \node [shape = circle, draw] (A) at (0,-5)   {$\bot$};
    \node [shape = circle, draw] (B) at (-5,0)   {$1$};
    \node [shape = circle, draw] (C) at (0,0)    {$2$};
    \node [shape = circle, draw] (D) at (5,0)    {$3$};
    \node [shape = circle, draw] (E) at (-5,5)   {$4$};
    \node [shape = circle, draw] (F) at (0,5)    {$5$};
    \node [shape = circle, draw] (G) at (0,10)   {$\top$};
  
    \draw[gray] (A) to node {} (B);
    \draw[gray] (A) to node {} (C);
    \draw[gray] (A) to node {} (D);
    \draw[gray] (B) to node {} (F);
    \draw[gray] (C) to node {} (F);
    \draw[gray] (D) to node {} (F);
    \draw[gray] (E) to node {} (G);
    \draw[gray] (F) to node {} (G);
    \draw[gray] (B) to node {} (E);
  
    \draw [loop below,blue,->,dotted,thick]          (A) to node {} (A);
    \draw [bend right=41,blue,->,dotted,thick]       (B) to node {} (A);
    \draw [blue,->,bend left=41,dotted,thick]        (C) to node {} (E);
    \draw [blue,->,bend left=60,dotted,thick]        (D) to node {} (E);
    \draw [bend right=60,blue,->,dotted,thick]       (E) to node {} (A);
    \draw [bend right,blue,->,dotted,thick]          (F) to node {} (E);
    \draw [bend right,blue,->,dotted,thick]          (G) to node {} (E);

    \draw [loop right,red,->,thick]                  (A) to node {} (A);
    \draw [bend right=45,red,->,thick]               (B) to node {} (D);
    \draw [red,->,loop right,thick]                  (C) to node {} (C);
    \draw [red,->,bend right,thick]                  (D) to node {} (F);
    \draw [bend right,red,->,thick]                  (E) to node {} (F);
    \draw [loop right,red,->,thick]                  (F) to node {} (F);
    \draw [bend left,red,->,thick]                   (G) to node {} (F);
  \end{tikzpicture}
\end{subfigure}
~
\begin{subfigure}[b]{0.45\textwidth}
  \centering
  \begin{tikzpicture}[scale=0.3,>=stealth]
    \tikzstyle{every node}=[font=\scriptsize]
    \node [shape = circle, draw] (A) at (0,-5)   {$\bot$};
    \node [shape = circle, draw] (B) at (-5,0)   {$1$};
    \node [shape = circle, draw] (C) at (0,0)    {$2$};
    \node [shape = circle, draw] (D) at (5,0)    {$3$};
    \node [shape = circle, draw] (E) at (-5,5)   {$4$};
    \node [shape = circle, draw] (F) at (0,5)    {$5$};
    \node [shape = circle, draw] (G) at (0,10)   {$\top$};
  
    \draw[gray] (A) to node {} (B);
    \draw[gray] (A) to node {} (C);
    \draw[gray] (A) to node {} (D);
    \draw[gray] (B) to node {} (F);
    \draw[gray] (C) to node {} (F);
    \draw[gray] (D) to node {} (F);
    \draw[gray] (E) to node {} (G);
    \draw[gray] (F) to node {} (G);
    \draw[gray] (B) to node {} (E);
  
    \draw [teal,loop below,->,dashed,thick]           (A) to node {} (A);
    \draw [teal,bend right=41,->,dashed,thick]        (B) to node {} (A);
    \draw [teal,bend right,->,dashed,thick]           (C) to node {} (A);
    \draw [teal,bend left,->,dashed,thick]            (D) to node {} (A);
    \draw [teal,bend right=60,->,dashed,thick]        (E) to node {} (A);
    \draw [teal,bend left,->,dashed,thick]            (F) to node {} (A);
    \draw [teal,bend left,->,dashed,thick]            (G) to node {} (A);

    \draw [bend right,-Rays,thick]            (D) to node {} (B);
    \draw [bend right,-Rays,thick]            (F) to node {} (B);
    \draw [bend right=12,-Rays,thick]            (G) to node {} (B);
  \end{tikzpicture}
\end{subfigure}
\caption{Join-endomorphisms {\color{blue} $f : \dottedarrow$}, {\color{red} $g : \rightarrow$} and $h = f \meetp{\jhspace{\Lat}} g$ ({\color{teal} $h : \dasharrow$}) on modular lattice $\Lat$. $\mathsf{x}$-head arrows depict maps of the function $\dapprox(c) = f(c) \meet g(c)$ s.t. $\dapprox(c) \neq h(c)$.}
\label{fig:ex-modular-lat}
\end{figure}

\begin{example}
  Consider the modular lattice $\Lat$ and $f,g \in \jhspace{\Lat}$ as depicted in Figure~\ref{fig:ex-modular-lat}. We want to compute $h = f \meetp{\jhspace{\Lat}} g$ represented with dashed teal arrows. Recall that Algorithm~\ref{alg:quicksp2} initiates with the function $\dapprox(c) = f(c) \meet g(c)$ for all $c \in \Lat$ and finds $h$ by correcting $\dapprox$ when $\dapprox(c) \neq h(c)$. In Figure~\ref{fig:ex-modular-lat}, $\mathsf{x}$-head arrows show the maps such that $\dapprox(c) \neq h(c)$. Notice that to correct the mapping of $\top$, it suffices to assign $\dapprox(4) \join \dapprox(5)$ to $\dapprox(\top)$, instead of considering all the pairs $(c,d)$ such that $c \join d = \top$, as the algorithm does. 
\end{example}

The above example suggests that, in modular lattices, we can decide whether or not a function is a join-endomorphism by checking the distribution property in the covers of every element in the lattice. We shall formalize this intuition and propose an adaptation of Algorithm~\ref{alg:quicksp2} for this particular case.
We specifically shall show that for modular lattices, the algorithm should only consider pairs in the cover set, $\selfcovers{c}$, for every $c \in \Lat$ (see Definition~\ref{def:covers}).

\subsubsection{Join-Endomorphisms in Modular Lattices}
\label{ssec:jendo-mod}

For the sake of the presentation, we present the following definition.

\begin{definition}
\label{def:coversp}
Let $\Lat$ be a lattice and $f : \Lat \to \Lat$ be a function.
We say that $f$ {\em preserves \hl{01-relations}} if for every $a,b,c \in \Lat$ such that $a,b \in \selfcovers{c}$, $f(a \join b) = f(a) \join f(b)$ holds.
\end{definition}
``fgfg''
The following is a well-known property of modular lattices; the cornerstone of the characterization of join-endomorphisms in this section.

\begin{proposition}{\cite{gratzer-lattice-2011}}
\label{prop:iso}
Let $\Lat$ be a lattice. $\Lat$ is modular iff for every $a,b \in \Lat$, the maps 
$\psi_a : {[a \meet b, b]} \to {[a,a \join b]}$ and $\varphi_b : {[a, a \join b]} \to {[a \meet b, b]}$ defined as $\psi_a(c) = c \join a$ and $\varphi_b(c) = c \meet b$, are mutually inverse isomorphisms.
\end{proposition}

The following theorem states that a function is a lattice join-endomorphism if it is \hl{01-relations} preserving.

\begin{theorem}
\label{thm:covers}
Let $\Lat$ be a modular lattice and let $f:{\Lat} \to {\Lat}$ be a function. $f$ is a join-endomorphism iff $f$ is bottom and \hl{01-relations} preserving.
\end{theorem}

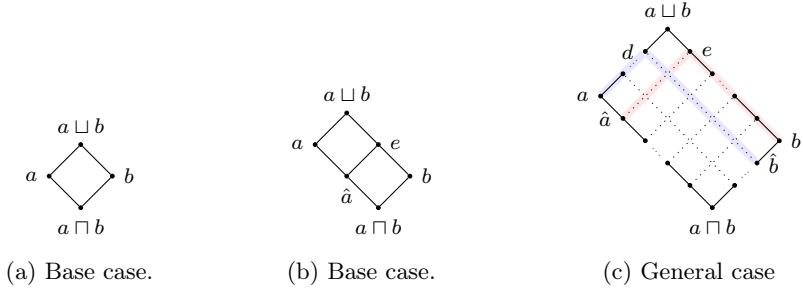
\begin{figure}
  \centering
  \begin{subfigure}[b]{0.28\textwidth}
    \centering
    \begin{tikzpicture}[scale=0.12,font=\footnotesize]
     \node[draw,circle,fill,label={below:$a \meet b$},inner sep=0.5pt] (22) at (0,-3.5) {};
     \node[draw,circle,fill,label={left: $a$},inner sep=0.5pt] (21) at (-3.5,0) {};
     \node[draw,circle,fill,label={right:$b$},inner sep=0.5pt] (12) at (3.5,0)  {};
     \node[draw,circle,fill,label={above:$a \join b$},inner sep=0.5pt] (11) at (0,3.5)  {};
     \draw (11) -- (12);
     \draw (11) -- (21);
     \draw (21) -- (22);
     \draw (12) -- (22);
    \end{tikzpicture}
    \caption{Base case.}
    \label{fig:mod-4}
  \end{subfigure}
  ~
  \begin{subfigure}[b]{0.28\textwidth}
  \centering
  \begin{tikzpicture}[scale=0.12,font=\footnotesize]
   \node[draw,circle,fill,label={below:$\hat{a}$},inner sep=0.5pt] (22) at (0,-3.5) {};
   \node[draw,circle,fill,label={left: $a$},inner sep=0.5pt] (21) at (-3.5,0) {};
   \node[draw,circle,fill,label={right:$e$},inner sep=0.5pt] (12) at (3.5,0)  {};
   \node[draw,circle,fill,label={above:$a \join b$},inner sep=0.5pt] (11) at (0,3.5)  {};
   \node[draw,circle,fill,label={below:$a \meet b$},inner sep=0.5pt] (31) at (3.5,-7) {};
   \node[draw,circle,fill,label={right:$b$},inner sep=0.5pt] (13) at (7,-3.5)  {};
  
   \draw (11) -- (12);
   \draw (11) -- (21);
   \draw (21) -- (22);
   \draw (12) -- (22);
   \draw (22) -- (31);
   \draw (12) -- (13);
   \draw (13) -- (31);
  \end{tikzpicture}
  \caption{Base case.}
  \label{fig:mod-6}
  \end{subfigure}
  ~
  \begin{subfigure}[b]{0.38\textwidth}
    \centering
    \begin{tikzpicture}[rotate=-135,scale=0.42,font=\footnotesize]
     \draw[dotted] (1.1,1.1) grid[step=1cm] (3.9,5.9);
     \foreach \x in {1,4}{
      \foreach \y in {1,2,...,6}{ 
       \node at (\x,\y) [draw,circle,inner sep=0.5pt,fill] (\x\y){};
       \ifthenelse{ \NOT \y = 4 \AND \NOT \y = 1}
       {\draw (\x,\y-1) -- (\x,\y)}{};
      }
     }
     \foreach \x in {2,3}{
      \foreach \y in {1,6}{ 
       \node at (\x,\y) [draw,circle,inner sep=0.5pt,fill] (\x\y){};
       \ifthenelse{\x = 2}{\draw (\x-1,\y) -- (\x,\y)}{};
       \ifthenelse{\x = 3}{\draw (\x,\y) -- (\x+1,\y)}{};
      }
     }
     \node also [label={above: $a \join b$}] (11);
     \node also [label={below: $a \meet b$}] (46);
     \node also [label={left:  $a$}]         (41);
     \node also [label={right: $b$}]         (16);
     \node also [label={right: $e$}]             (12);
     \node also [label={left:  $d$}]             (21);
     \node also [label={right: $\hat{b}$}]             (26);
     \node also [label={left:  $\hat{a}$}]             (42);
    
     \begin{scope}[on background layer]
      \draw [line width=3pt,red!10]  (12) -- (42);
      \draw [line width=3pt,red!10]  (12) -- (16);
      \draw [line width=3pt,blue!10] (21) -- (41);
      \draw [line width=3pt,blue!10] (21) -- (26);
     \end{scope}
     \begin{scope}[transform shape]
     \node at (1,3.6) {$\vdots$};
     \node at (4,3.6) {$\vdots$};
     \node at (2.55,1) {$\dots$};
     \node at (2.55,6) {$\dots$};
     \end{scope}
    \end{tikzpicture}
    \caption{General case}
    \label{fig:mod-nm}
    \end{subfigure}
  \caption{Isomorphism on modular lattices with different size of intervals}
\end{figure}

\begin{proof}
  The only-if direction is straightforward. For the if direction, let $f$ be a function such that $f(\bot) = \bot$ and, for every $a,b,c \in \Lat$ with $a,b \in \selfcovers{c}$, $f(a \join b) = f(a) \join f(b)$ holds.
  
  By Proposition~\ref{prop:monotonicity}, to prove that $f$ is a join-endomorphism, it suffices to show that $(\star)$ $f(a \join b) = f(a) \join f(b)$ for every $a,b \in \Lat$. 
  
  We define $n = \big|[a, a \join b]\big| = \big|[a \meet b, b]\big|$ and $m = \big|[b, a \join b]\big| = \big|[a \meet b, a]\big|$ where $a,b \in \Lat$ are any pair of elements such that $a \| b$. To prove $(\star)$ we proceed by induction on the size of those intervals, i.e. on both $n$ and $m$. 
  
  For the base cases, if $n = m = 2$ (Figure~\ref{fig:mod-4}) the proof follows by the assumption on $f$. Without loss of generality, assume $n = 2$ and $m = 3$ as shown in Figure~\ref{fig:mod-6}. Since $f$ preserves \hl{01-relations}, we have $f(a \join b) = f(a) \join f(e)$ and $f(e) = f(\hat{a}) \join f(b)$ where $b \cl e \lcov a \join b$ and $a \meet b \cl \hat{a} \lcov a$. Then, we obtain $f(a \join b) = f(a) \join f(\hat{a}) \join f(b) = f(a \join \hat{a}) \join f(b) = f(a) \join f(b)$ as wanted.

  For the inductive case, assume that $(\star)$ holds for every $2 < n' < n$ and every $2 < m' < m$, i.e. for any intervals $[a,d]$ and $[b,e]$ of size $n'$ and $m'$, respectively, where $a \cl d \cl a \join b$ and $b \cl e \cl a \join b$. By assumption on $f$, $f(a \join b) = f(d) \join f(e)$ with $a \cl d \lcov a \join b$ and $b \cl e \lcov a \join b$ as shown in Figure~\ref{fig:mod-nm}. From the induction hypothesis and Proposition~\ref{prop:iso}, we know that $f(d) = f(a) \join f(\hat{b})$ and $f(e) = f(\hat{a}) \join f(b)$ where $\hat{a} \lcov a$ and $\hat{b} \lcov b$. Thus, the proof follows from the fact that $f$ preserves \hl{01-relations}: $f(a \join b) = f(a \join \hat{a}) \join f(b \join \hat{b}) = f(a) \join f(b)$.
  
  If $a$ and $b$ are comparable, let us say $a \cleq b$ (the other case is analogous), the proof goes as before with $\hat{b} = d$ and $e = b$.
\end{proof}

Now we have the means to present Algorithm~\ref{alg:quicksp2} adapted to modular lattices.

\subsubsection{Algorithm for Modular Lattices}
\label{ssec:algo-modular}

Recall the sets $\Cseta{w}$, $\Fseta{w}$ and $\Sseta{w}$ used by the algorithm to separate the pairs $(u,v) \in \Lat^2$: Namely, with $w = u \join v$, $(u,v) \in \Sseta{w}$ if $\dapprox(u) \join \dapprox(v) = \dapprox(w)$, $(u,v) \in \Cseta{w}$ if $\dapprox(u) \join \dapprox(v) \cl \dapprox(w)$ and, $(u,v) \in \Fseta{w}$ otherwise. We shall show that for any modular lattice $\Lat$, we can construct these sets by using elements in the sets $\selfcovers{w} = \covers{w} \cup \{w\}$ for every $w \in \Lat$. Roughly speaking, we consider only pairs of elements $(u,v)$ that are covered by an element $w$.

\begin{observation}
\label{obs:algo-gmeet-covers}
Let $\Lat$ be a modular lattice, $S \subseteq \jhspace{\Lat}$ and $W = \bigcup_{w \in \Lat} \selfcovers{w} \subseteq \Lat$. If we form the sets $\Cset$, $\Fset$ and $\Sset$ with pairs of elements in $W$, it is an immediate consequence of Theorem~\ref{theorem:approx-correct} that Algorithm~\ref{alg:quicksp2} terminates. Therefore, we have found a function $\dapprox'$ (see Theorem~\ref{theorem:approx-correct}) such that $\dapprox' = \Dfun{S}$. Moreover, Theorem~\ref{thm:covers} guarantees that such a function is a join-endomorphism. 
\end{observation}

\subsubsection*{Experimental Results}
\label{ssec:exp-covers-alg}

Figure shows\dots

\paragraph{Complexity for Modular Lattices}
As we have shown, the time complexity of Algorithm~\ref{alg:quicksp2} reduces drastically with the adaptation for modular lattices. Specifically, the reduction is due to the initialization step (line 2 in Algorithm~\ref{alg:quicksp2}) of the sets $\Cset$, $\Fset$ and, $\Sset$. Since we are taking elements in the set $W \subseteq \Lat$, we reduce the number of pairs that the algorithm loops through. However, if the lattice is $\M_n$, the set $\selfcovers{\top} = \{\top, 1, \ldots, n\}$; in other words, the algorithm will consider all the elements of $\M_n$. This implies that the algorithm takes no advantage of the modularity of the lattice and the worst-time time complexity is still $O(mn + n^3)$.

We conclude this section with a small example where join-endomorphisms represent dilation operators from Mathematical Morphology~\cite{bloch-mm-2007}. We use the algorithms presented above to compute  the greatest dilation below a given set of dilations and illustrate its result for a simple image. 

\subsection{A Mathematical Morphology Example}
\label{ssec:ex-mm}
Mathematical morphology (MM) is a theory, based on topological, lattice-theoretical  and geometric concepts, for the analysis of geometric structures. Its algebraic framework comprises \cite{bloch-mm-2007,ronse-mmclat-1990,stell-mmquantales-2009}, among others, complete lattices together with certain kinds of morphisms, such as \emph{dilations},
defined as \emph{join-endomorphisms} \cite{ronse-mmclat-1990}. 
Our results give  bounds about the number of all dilations over certain specific finite lattices and also efficient algorithms to compute their infima.

A typical application of MM is image processing.
Consider the space $G = \mathbb{Z}^2$. A dilation~\cite{bloch-mm-2007} by
$s_i \subseteq \pcal(G)$ is a function $\func{\delta_{s_i}}{\pcal(G)}{\pcal(G)}$ such that $\delta_{s_i}(X)  = \{x+e\ |\ x \in X \text{ and } e \in s_i\}$. The dilation $\delta_{s_i}(X)$ describes the interaction of an image $X$ with the \emph{structuring element} $s_i$.
Intuitively, the dilation of $X$ by $s_i$ is the result of superimpose $s_i$ on every activated pixel of $X$, with the center of $s_i$ aligned with the corresponding pixel of $X$. Then, each pixel of every superimposed $s_i$ is included in $\delta_{s_i}(X)$.

\begin{figure}
 \centering
 \includegraphics[scale=0.5]{\mainDir/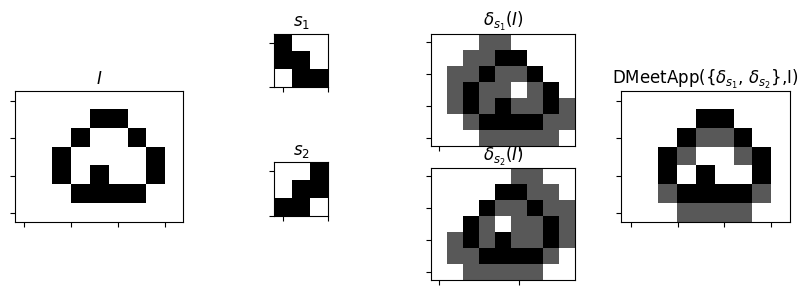}
 \caption{\footnotesize{Binary image $I$ (on the left). Dilations $\delta_{s_1}$, $\delta_{s_2}$ for structuring elements $s_1$, $s_2$. On the right $(\delta_{s_1} \meetp{\jhspace{\Lat}} \delta_{s_2})(I)$. New pixels of the image after each operation in grey and black.}}
 \label{fig:mathmorph}
\end{figure}

Let $\Lat$ be the powerset lattice for some finite set $D\subseteq G.$
It turns out that the dilation $\Dfun{S}$ corresponds to the intersection of the structuring elements of the corresponding dilations in $S$.
Figure~\ref{fig:mathmorph} illustrates $\Dfun{S}$ for the two given dilations
$\delta_{s_1}(I)$ and $\delta_{s_2}(I)$ with structuring elements $s_1$ and
$s_2$ over the given image $I$.


\subsection{Proof of Theorem~\ref{theorem:approx-correct}}
\label{sec:proof-camilo-thm}

Let $L$ be a finite  lattice, $u,v\in \Lat$, $\dapprox: \Con \rightarrow \Con$ and $S \subseteq \sfunspace{L}$. Assume  $\dapprox \cgeq \Dfun{S}$ holds, and consider the following updates:
\begin{enumerate}
\item when
$\dapprox(u) \join\dapprox(v)\cl\dapprox(u\join v)$, assign
$\dapprox(u\join v) \leftarrow \dapprox(u)\join\dapprox(v)$
\item when
$\dapprox(u)\join\dapprox(v)\not\cleq\dapprox(u\join v)$, assign
$\dapprox(u)\gets \dapprox(u)\meet\dapprox(u\join v)$ and also 
$\dapprox(v)\gets \dapprox(v)\meet\dapprox(u\join v)$
\end{enumerate}
Let $\dapprox'$ be the function resulting after the update. Then, (1) $\dapprox' \cl \dapprox$ and (2) $\dapprox' \cgeq \Dfun{S}$

\begin{proof}
Let $L$ be a finite  lattice, $S \subseteq \sfunspace{L}$, and $h = \Dfun{S}$.

For update (1), given the condition, the assignment obviously decreases $\dapprox(u \join v)$, so $\dapprox' \cl \dapprox$. For the invariant, since $\dapprox \cgeq h$, then, $\dapprox(u) \cgeq h(u)$ and $\dapprox(v) \cgeq h(v)$, and therefore, $\dapprox'(u\join v)=\dapprox(u) \join \dapprox(v) \cgeq h(u)\join h(v) = h(u \join v)$.
  
For update (2), the assignments either decrease $\dapprox(u)$ or $\dapprox(v)$ (or both). To see why, assume the opposite, $\dapprox(u) = \dapprox(u)\meet\dapprox(u\join v) \imp \dapprox(u) \cleq \dapprox(u\join v) $,  and also $\dapprox(v) = \dapprox(v)\meet\dapprox(u\join v) \imp \dapprox(v) \cleq \dapprox(u\join v)$. Therefore, $  \dapprox(u)\join\dapprox(v) \cleq \dapprox(u\join v)$, contradicting the condition for update 2. Assignments in update 2 also preserve the invariant $\dapprox \cgeq h$.

Assume $\dapprox(u)\meet\dapprox(u\join v)\cl \dapprox(u)$ (otherwise the invariant holds trivially). By the invariant hypothesis for $\dapprox$ before the assignment, we have that $ \dapprox(u) \cgeq h(u)$ and $ \dapprox(u\join v) \cgeq h(u \join v)$. Therefore, 
\begin{align*}
  \dapprox(u) \meet \dapprox(u\join v)
& \cgeq h(u) \meet h(u \join v)\\
& = h(u) \meet (h(u) \join h(v))\\
& = h(u) \join (h(u) \meet h(v))\\
& = h(u)
\end{align*}
The proof for $\dapprox'(v)$ is analogous.
\end{proof}

\section{Conclusions and Related Work}
\label{sec:related}

We have shown that given a lattice $\Lat$ of size $n$ and a set $S\subseteq \sfunspace{\Lat}$ of size $m$, $\Dfun{S}$ can be computed in the worst-case 
in $O(n+ m{\log n})$ binary lattice operations for powerset lattices, $O(mn^2)$ for lattices of sets, and $O(nm + n^3)$ for arbitrary lattices. We illustrated the experimental performance of our algorithms and a small example from mathematical morphology. 
 
 In \cite{HABIB1996391} a bit-vector representation of a lattice is discussed. This work gives algorithms of   logarithmic (in the size of the lattice) complexity for join and meet operations. These results count bit-vector operations.  From \cite{birkhoff-lattice-1940} we know that $\sfunspace{\Lat}$ is isomorphic to the downset of $(P \times P^{\it op})$, where $P$ is the set of join-prime elements of $L$, and that this, in turn, is isomorphic to the set of order-preserving functions from $(P\times P^{\it op})$ to $\mathbf{2}.$ Therefore, for the problem of computing $\Dfun{S}$, we get bounds $O(m \log_2(2^{(n^2)}) = O(mn^2)$ for set lattices and $O(m(\log_2n)^2)$  for powerset lattices where $n=|\Lat|$ and $m = |S|.$ This, however, assumes a bit-vector representation of a lattice isomorphic to $\sfunspace{\Lat}$. Computing this representation takes time and space proportional to the size of $\sfunspace{\Lat}$ \cite{HABIB1996391} which could be exponential as stated in the present paper. Notice that in our algorithms the input lattice is $\Lat$ instead of $\sfunspace{\Lat}.$ 
 
We have stated the cardinality of the set of join-endomorphisms $\sfunspace{\Lat}$  for significant families of lattices. To the best of our knowledge we are the first to establish the cardinality $(n+1)^2 + n!\laguerre{n}{-1}$ for the lattice $\M_n.$  The cardinalities $n^{\log_2 n}$ for power sets (boolean algebras) and  $\binom{2n}{n}$ for linear orders can also be found in the lattice literature \cite{birkhoff-lattice-1940,jipsen-ramics-2017,santocanale-words-2019}. Our original proofs for these statements can be found in the technical report of this paper \cite{quintero:hal-02422624}. 
 
The lattice $\sfunspace{\Lat}$ have been studied in \cite{gratzer-latjoinend-1958}. The authors showed that a finite lattice $\Lat$ is distributive iff  $\sfunspace{\Lat}$ is distributive.  A lower bound of $2^{2n/3}$ for the number of monotonic self-maps of any finite poset $L$ is given in \cite{Duffus1992}. Nevertheless to the best of our knowledge, no other authors have studied the problem of determining the size $\sfunspace{\Lat}$ nor algorithms for computing  $\Dfun{S}.$ We believe that these problems are important, as argued in the Introduction; algebraic structures consisting of a lattice and join-endomorphisms are very common in mathematics and computer science. In fact, our interest in this subject arose in the algebraic setting of spatial  and epistemic constraint systems \cite{GuzmanKQRRV19} where continuous join-endomorphisms, called space functions, represent knowledge and the infima of endomorphisms correspond to distributed knowledge. We showed in \cite{GuzmanKQRRV19}  that  distributed knowledge can be computed in $O(mn^{1+\log_2(m)})$ for distributive lattices and $O(n^4)$ in general. In this paper we have provided much lower complexity orders for computing infima of join-endomorphisms.  Furthermore \cite{GuzmanKQRRV19} does not provide the exact cardinality of the set of space functions of a given lattice.

As future work we plan to explore in detail the applications of our work in mathematical morphology and computer music \cite{Rueda2004}. Furthermore, in the same spirit of \cite{jipsen-genlattice-2015} we have developed algorithms to generate distributive and arbitrary lattices. In our experiments, we observed that for every lattice $\Lat$ of size $n$ we generated, $n^{\log_2 n}\leq |\sfunspace{\Lat}| \leq (n+1)^2 + n!\laguerre{n}{-1}$ and  if the generated lattice was distributive,  $n^{\log_2 n}\leq |\sfunspace{\Lat}| \leq \binom{2n}{n}.$ We plan to establish if these inequalities hold for every finite lattice.

\subsubsection*{Acknowledgments.} We are indebted to the anonymous 
referees and editors of RAMiCS 2020 for helping us to improve one of the complexity bounds, some proofs, and the overall quality of the paper.

\bibliographystyle{elsarticle-num}
\bibliography{bibliography/biblio}



\end{document}